\documentclass[a4paper,11pt]{article}
\usepackage{jheppub}

\pdfoutput=1
\usepackage{hyperref}
\usepackage{graphicx}
\usepackage[utf8]{inputenc}
\usepackage{xspace}
\usepackage{amsmath,amssymb}
\usepackage{bbm}
\usepackage{xcolor}
\usepackage{booktabs}
\usepackage{subcaption}
\usepackage{mathtools}
\usepackage[ruled,norelsize]{algorithm2e}
\usepackage{empheq}
\usepackage{cancel}
\usepackage{multirow}
\usepackage{slashed}
\usepackage{feynmp}
\usepackage{feynmp-auto}
\usepackage{tikz}
\usetikzlibrary{decorations.pathmorphing}
\usetikzlibrary{decorations.markings}
\usetikzlibrary{positioning, shapes, snakes, arrows}
\tikzset{
	quark/.style={postaction={decorate},
		decoration={markings,mark=at position .5 with {\arrow[#1]{latex}}}},
	scalar/.style={dashed,postaction={decorate},
		decoration={markings,mark=at position .5 with {\arrow[#1]{latex}}}},
	gluon/.style={decorate,
		decoration={coil,amplitude=2pt, segment length=2pt,  pre length=.1cm, post length=.1cm}},
	boson/.style={-latex,decorate, decoration={snake, segment length=4pt, amplitude=1.8pt, pre length=.1cm, post length=.25cm}},
	photon/.style={decorate, decoration={snake, segment length=4pt, amplitude=1.8pt,  pre length=.1cm, post length=.1cm}},
	dphoton/.style={decorate, decoration={snake, segment length=4pt, amplitude=1.8pt,  pre length=.1cm, post length=.25cm},-latex}
}

\DeclareFontFamily{U}{rcjhbltx}{}
\DeclareFontShape{U}{rcjhbltx}{m}{n}{<->rcjhbltx}{}

\newcommand{\pqbar}{p_{\bar{q}}}
\newcommand{\pq}{p_q}
\newcommand{\tildepq}{\tilde{p}_q}
\newcommand{\tildepqbar}{\tilde{p}_{\bar{q}}}

\title{Exploring soft anomalous dimensions for $1/Q$ power corrections}

\author{Mrinal Dasgupta and Farid Hounat}

\affiliation{Department of Physics \& Astronomy, University of
  Manchester, Manchester M13 9PL, United Kingdom}

\emailAdd{mrinal.dasgupta@manchester.ac.uk}
\emailAdd{farid.hounat@manchester.ac.uk}

\abstract{In this article we study, via analytical methods, $1/Q$ non-perturbative power corrections to event shape mean values, addressing in particular the question of their interplay with soft perturbative emissions. Specifically we point out that energy-ordered soft perturbative emissions that precede a non-perturbative emission, give rise to terms of the form $\frac{1}{Q} \left (\alpha_s  \ln \frac{Q}{\Lambda} \right)^n$. While such terms are formally higher order in the strong coupling, their form suggests that they can numerically compete with the leading $1/Q$ term while also modifying the $Q$ dependence of the result. The resummation of such power-suppressed but logarithmically enhanced terms lends an anomalous dimension to the  leading $1/Q$ power correction.  In order to argue for the presence of such an anomalous dimension, we formulate a method to compute the first order in $\alpha_s$ correction for the mean values of the thrust $1-T$ and $C$-parameter observables. We comment on our findings in light of the standard picture of universality of $1/Q$ power corrections for event shape variables and implications for phenomenology.}

\begin{document}
\setlength{\parskip}{0pt}
\maketitle
\flushbottom

\section{Introduction}
The issue of non-perturbative corrections has for long been a thorn in the side of achieving high precision across the full range of phenomenological QCD studies at colliders. With the current status of experimental precision in collider phenomenology and the potential wealth of data that might be forthcoming from possible future colliders such as FCC-ee \cite{ref}, the focus on precise theoretical estimates has never been stronger. This need for precision poses a challenge which is being met on several fronts. Examples include development of fixed-order studies at NNLO \cite{Gehrmann-DeRidder:2007nzq, Dissertori:2008cn, Gehrmann-DeRidder:2009fgd,Weinzierl:2008iv, Weinzierl_2009yz,Gehrmann:2019hwf,Alvarez:2023fhi} and $\mathrm{N^3LO}$ \cite{Ebert:2020sfi} for event shape variables, as well as resummed calculations at NNLL \cite{Bozzi:2005wk,Bozzi:2010xn, Stewart_2011, Banfi:2011dm, Banfi:2012jm, Banfi:2015pju, Jouttenus_2013,Kang:2013nha,Banfi:2014sua,Becher:2015gsa, Banfi:2016zlc,Becher_2016, Tulipant:2017ybb, Banfi:2018mcq,Gao:2019ojf,Kardos:2020gty,Dasgupta:2022fim,vanBeekveld:2023lsa,Chen:2023zlx,Bhattacharya:2023qet} and $\mathrm{N^3LL}$  accuracy \cite{Becher:2008cf, Chien:2010kc,Abbate:2010xh,Hoang:2014wka} for observables in various collider processes. There has also been considerable recent progress on parton shower accuracy via achievement of general NLL accuracy \cite{Forshaw:2019ver, Forshaw:2020wrq,Dasgupta:2020fwr,Hamilton:2020rcu,Nagy:2020dvz,Nagy:2020rmk,Karlberg:2021kwr,Hamilton:2021dyz,vanBeekveld:2022zhl,vanBeekveld:2022ukn,Herren:2022jej,vanBeekveld:2023chs,Preuss:2024vyu,Hoche:2024dee} as well as NNLL accuracy for $e^{+}e^{-}$ event shape variables \cite{FerrarioRavasio:2023kyg,vanBeekveld:2024wws}.

While systematic progress on changing the state of the art of perturbative aspects of QCD is challenging conceptually and computationally, it is at least feasible to work to a well defined agenda aimed at capturing higher accuracy perturbative terms of various kinds. However even such well defined perturbative programs would, if pushed very far, ultimately run into issues related to the lack of convergence of the perturbative expansion via factorial infrared (IR) renormalon divergences, which have been argued to relate to the ultimate presence of non-perturbative effects (see Ref.~\cite{Beneke:1998ui} for a review). Here due to lack of direct first principles insight into the confinement domain, it is only possible to resort to models whose parameters can be constrained using experimental data, and the approach to systematic theoretical improvement is often unclear.

Amongst models for non-perturbative physics, perhaps most widely used for studies of non-perturbative corrections are hadronization models in Monte Carlo event generators (see Ref.~\cite{Campbell:2022qmc} and references therein). While these have proved invaluable given their general applicability, they also have some well known limitations. One such issue is the presence of a number of adjustable parameters which do not have a clear connection to QCD first principles and hence might mask erroneous physical assumptions \cite{Webber:1997zj}. Another related issue is the fact that the tuning of hadronization model parameters corrects the typically limited accuracy parton level results of event generators to hadron level data, which may result in incorrectly attributing perturbative mis-modelling to non-perturbative hadronization. Somewhat more limited but still widely studied are analytical models of power corrections, which have the potential advantage of being inspired by direct connections to theoretical aspects such as renormalons, and are cleaner in that they involve the introduction of only a limited number of non-perturbative parameters. Analytical models that have traditionally been used in studies of $1/Q$ power corrections include those based on a universal infrared finite coupling \cite{Dokshitzer_1995,Dokshitzer_1996,Dokshitzer:1997ew, Dokshitzer_1998,Dokshitzer_1998_2, Dokshitzer:1998qp,Banfi:2000si, Banfi:2000ut, Banfi:2001sp, Banfi:2001pb} and those based on non-perturbative shape functions \cite{Korchemsky:1999kt, Korchemsky:2000kp, Bauer:2003di,Lee:2006nr, Abbate:2010xh, Hoang:2014wka, Hoang:2015hka}.

Particularly in the context of analytical models of non-perturbative effects, a class of observables that has received substantial attention from a phenomenological viewpoint are event shape variables. On the one hand such observables are classic examples of IRC safe observables that, as already mentioned, lend themselves to precise predictions in QCD perturbation theory at fixed-order as well as resummed calculations to all orders. For this reason they have been used for tests of QCD, for instance via extractions of the strong coupling and studies of its running behaviour \cite{Jones:2003yv,Pahl:2009aa,Davison:2009wzs,Gehrmann:2010uax,Verbytskyi:2019zhh,Bethke:2009ehn,Dissertori:2009ik,OPAL:2011aa,Kardos:2018kqj,dEnterria:2022hzv}. At the same time several event shapes are affected by significant $1/Q$ linear power corrections which numerically compete with higher order perturbative corrections and have a pronounced impact on the quality of theory-experiment comparisons, limiting the accuracy of $\alpha_s$ extractions (see Ref.~\cite{Nason:2023asn} for examples of most recent studies). However event shapes prove also to be good observables to study non-perturbative effects in a phenomenological context. Perhaps the most widely used analytical model is the universal IR finite coupling approach pioneered by Dokshitzer and Webber (DW) \cite{Dokshitzer_1995,Dokshitzer:1997ew}. This idea, with some theoretical improvements, has stood the test of time and it has for long been standard practice to carry out studies of event shapes with the aim of extracting $\alpha_s$ and $\alpha_0$, where the latter is related to a moment of the assumed infrared finite modification to the perturbative coupling. 

One significant advantage of the DW model is its simplicity and its connection to an underlying physical picture. It generates the non-perturbative correction via an emission of a ``non-perturbative gluon" or ``gluer'' (we refer the interested reader to Ref.~\cite{Dokshitzer:1999ai} where this terminology is explained more vividly), with transverse momentum $k_t \sim \Lambda$, with $\Lambda$ being the QCD scale. Such emissions have a strength given by an assumed universal non-perturbative coupling, while the standard perturbative coupling would instead diverge. Computing how the gluer emission modifies the event shape is then straightforward, i.e. proceeds like a regular perturbative calculation, and leads to predictions for power corrections involving only a single non-perturbative parameter $\alpha_0$, related to the average coupling strength over the non-perturbative region. 

Another attractive feature of the DW model and related approaches is that it proves possible to some extent to systematically improve the model from a theoretical viewpoint. An important past step in this direction has been the calculation of two-loop enhancement factors (``Milan factors") \cite{Dokshitzer_1998,Dokshitzer_1998_2, Dasgupta:1998xt,Dasgupta:1999mb} to account for the impact of gluer decay on event shape observables \cite{Nason:1995np}. Much more recently there has also been significant effort devoted to computing power corrections in the three-jet limit of event shape distributions \cite{Luisoni:2020efy,Caola:2021kzt,Caola:2022vea,Nason:2023asn}. These calculations were driven by the observation that $\alpha_s$ extractions in some well-known event shape distributions were being affected by an erroneous assumption that the non-perturbative shift of the distribution could simply be approximated by the shift computed in the two-jet limit ($v \to 0$ for some two-jet event shape $v$) across the full spectrum, i.e. for any value of $v$ \cite{Luisoni:2020efy}. As a result it has recently become possible to estimate the non-perturbative shift to event shape distributions as a function of the shape variable $v$. 

In this paper we wish to address another element which has so far been missing from discussions and analytical studies of non-perturbative power corrections, but which is necessary to further improve the existing theoretical models. This concerns the issue of the {\it{perturbative evolution}} of the $1/Q$ non-perturbative corrections and the consequent anomalous dimension associated to these power corrections. To appreciate the presence of an anomalous dimension, one has to think about the interplay of a non-perturbative gluer emission with soft but still perturbative gluons i.e. those with $Q \gg k_t \gg \Lambda$. As we shall show, gluer emission in the presence of such soft perturbative emissions leads to a series of terms that are power suppressed but logarithmically enhanced in $Q$, i.e of the form $\frac{\Lambda}{Q} \left(\alpha_s \ln \frac{Q}{\Lambda} \right)^n$ where $n$ is the number of soft perturbative emissions considered. Given that $\alpha_s(Q) \propto 1/\ln(Q/\Lambda)$ these terms can numerically compete with the leading $1/Q$ correction and should evidently be resummed. This resummation, which is precisely of the same nature as that needed for leading non-global logarithms \cite{Dasgupta:2001sh,Dasgupta:2002bw}, will effectively yield an anomalous dimension which governs the perturbative $Q$ dependence of the leading $1/Q$ power correction. \footnote{The anomalous dimension we discuss here is physically distinct from other studies in the literature which examine the $Q$ dependence of non-perturbative effects such as the anomalous dimension associated to hadron mass effects \cite{Salam:2001bd} and recent studies for energy correlators \cite{Lee:2024esz,Chen:2024nyc}.}

There has only been a very limited discussion of the anomalous dimension issue in the literature so far. A brief mention of the likely presence of such terms can be found in Ref.~\cite{Dasgupta:2003iq} where an absence of any concrete calculations to support the idea was also noted. More recently Ref.~\cite{Luisoni:2020efy} dedicated to studies of the non-perturbative shift $\zeta(C)$ for the $C$-parameter distribution also mentions the presence of the anomalous dimension, and the need to include it in order to improve the fixed-order studies carried out therein. In the context of that paper, this would imply including contributions from soft energy-ordered gluons between the transverse momentum scale $k_t \sim QC$ of the perturbative emission that sets the value of the $C$-parameter, and the non-perturbative gluon with $k_t \sim \Lambda$.  \footnote{We also note that Ref.~\cite{Nason:1995np} mentions a possibility that $1/Q$ corrections may develop $\ln \frac{Q}{\Lambda}$ logarithmic enhancements at higher orders.}

In the current paper we present the first dedicated analytical calculation for the effect in question, which we hope will help to elucidate its origin and its nature. Our calculation shares a number of similarities to recent calculations devoted to the three-jet limit for event shapes \cite{Luisoni:2020efy,Caola:2021kzt,Caola:2022vea,Nason:2023asn}, but is distinct in that it focuses on a soft perturbative gluon which dresses the two hard parton $q\bar{q}$ system,  which lets us isolate the anomalous dimension. Furthermore, for simplicity, we focus on analytically computing the leading-order term in the perturbative series of soft emissions i.e. we compute a term of the form $\frac{\Lambda}{Q} \alpha_s \ln \frac{Q}{\Lambda}$.  We also confine our attention to the case of event shape mean values while studying two common shape variables, the thrust and the $C$-parameter.  Our results enable us to shed light on the anomalous dimension and on how it impinges on the key question of universality of $1/Q$ power corrections to linear event shape variables.

This paper is organised as follows: in section \ref{sec:gen} we make some general remarks and set up the steps needed for the calculations that follow. Section \ref{sec:thrust} is devoted to a calculation of the non-perturbative correction to the mean value for the thrust $\langle 1-T \rangle$. Here we discuss how one may compute the change in thrust induced by the emission of a gluer, in the presence of a soft perturbative emission. This requires us to choose a recoil scheme and we discuss our specific choice while also commenting on the potential dependence on recoil scheme choices. We then consider the emission of a gluer from a $q\bar{q} g$ configuration involving a soft perturbative gluon. The emission pattern can be cast in terms of three emitting dipoles and we compute the change in thrust for gluer emission from each dipole. Next we carry out integrals over the dipole emission phase space which give us the logarithmically enhanced non-perturbative corrections we seek. Section \ref{sec:c} is devoted to similar steps for the $C$ parameter. Having obtained results for non-perturbative corrections to $\langle1-T\rangle$ and $\langle C \rangle$, we compare the results and comment on the universality of the anomalous dimensions. Finally we conclude in section \ref{sec:disc} with a short summary of our findings and a discussion of potential future directions.

\section{General idea and setup for calculations}
\label{sec:gen}
\begin{figure}
\begin{center}
\includegraphics{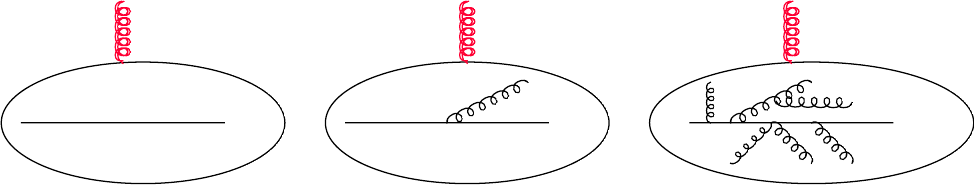}
\caption{Schematic representation of gluer emission from a) a hard $q\bar{q}$ dipole (left) b) a $q\bar{q}$ dipole dressed with an additional soft gluon with transverse momentum $Q \gg k_t \gg \Lambda$ (centre), an ensemble consisting of a hard $q\bar{q}$ dipole dressed by soft gluons strongly ordered in transverse momenta such that $Q \gg k_{t1} \gg k_{t2} \cdots \gg \Lambda$ (right).  }
\label{fig:gluons}
\end{center}
\end{figure}

As already stated, we revisit the derivation of $1/Q$ power corrections in the Dokshitzer-Webber (DW) model \cite{Dokshitzer_1995} to take account of higher orders in $\alpha_s$ i.e. an interplay between non-perturbative and perturbative gluon radiation. We are interested in a special limit, which as we will show will produce a leading-order term of the form $\frac{\Lambda}{Q} \alpha_s \ln \frac{Q}{\Lambda}$. To physically motivate the presence of such terms we first consider the standard DW model where the leading $1/Q$ power correction is triggered by the emission of a gluer with $k_t \sim \Lambda$, from a hard $q \bar{q}$ system, as shown on the leftmost picture in Fig.~\ref{fig:gluons}. At higher orders however one has to consider more complicated emitting ensembles and in particular the evolution of the emitting perturbative system via soft gluon emission. When studying order $\alpha_s$ power corrections, for example, we need to account for the presence of a single perturbative gluon so that the gluer now comes from a $q \bar{q} g$ system. While the gluer will be responsible for modifying a given observable at the $1/Q$ power suppressed level as usual, the emission of the additional perturbative gluon brings a factor of $\alpha_s$, which should, naively speaking, relegate this configuration to a higher-order effect beyond the accuracy of the DW model. However, if one correctly accounts for the fact that the emission of the perturbative gluon can in fact be soft-enhanced and strongly ordered in energy (or equivalently for our purposes in transverse momenum $k_t$) wrt the softer gluer, the perturbative emission can have a range of transverse momenta $Q \gg k_t \gg \Lambda$. In this situation the integrated emission probability for the perturbative gluon emission produces terms varying as $\sim \alpha_s \ln \frac{Q}{\Lambda}$. Where such terms do not cancel against virtual corrections, we may therefore expect a logarithmically enhanced power correction to emerge from this double-soft energy-ordered region. 

The above observations easily generalise to an ensemble of perturbative soft, commensurate angle, but energy-ordered gluons as shown in the rightmost picture of Fig.~\ref{fig:gluons}.
 This type of configuration can produce a power correction of the form $ \left(\alpha_s \ln \frac{Q}{\Lambda} \right)^n \frac{\Lambda}{Q}$. A resummation of the large logarithms would then be required in order to accurately capture the effective $1/Q$ power correction. Moreover the coefficient of the $1/Q$ term acquires a further $Q$ dependence i.e. the anomalous dimension for the $1/Q$ correction. Such soft commensurate-angle energy-ordered gluon ensembles ought to be familiar in the context of non-global (NG) observables. The physics we are highlighting here is essentially the same except that the softest gluon, which in the NG case is tagged via the observable, is now treated as a non-perturbative gluer and triggers a $1/Q$ correction. The perturbative ensemble instead generates logarithmic enhancements which are physically related to those seen in the NG case. In both the present case and the NG case, it is the fact that we are sensitive to the contribution of the softest gluon in the ensemble, which triggers the contribution from soft energy-ordered gluons at commensurate angles. In the present case the sensitivity is because we are associating the hadronization correction to the softest gluon with $k_t \sim \Lambda$, while in the NG case we are sensitive to the softest emission by virtue of measuring in only a limited phase-space region.
  \footnote{We note here that where the gluer is  emitted at disparate angles to the perturbative emission, i.e. independently, there will be no contribution to the anomalous dimension which will be discussed in more detail later in the article.}
 
Below we describe in more detail the origin and calculation of the leading-order $\alpha_s \ln \frac{Q}{\Lambda}$ correction. We choose to illustrate our arguments on the relatively simple but classic case of event shape mean values in $e^{+}e^{-}$ annihilation, where we examine both the thrust $1-T$, and the $C$-parameter in detail.
 
In terms of our general strategy, we shall apply the DW model by determining the change in the event shape $\delta v$ induced by gluer emission, weighting it by the probability of gluer emission from the underlying perturbative system, and integrating over the available phase-space to compute the impact on the mean event shape. We will therefore evaluate an expression of the schematic form

\begin{equation}
\label{eq:schem}
 \langle \delta v \rangle^{\mathrm{NP}} = \frac{1}{\sigma} \int d\Gamma \, \mathcal{M}^2 \, \delta v,
 \end{equation}
where $\mathcal{M}^2$ is the matrix element for gluer emission and $d\Gamma$ the differential phase space, while $\delta v$ denotes the change in some event shape $v$ induced by gluer emission.

To compute $\delta v$ in the usual DW model one simply notes that the pre-emission value of the event shape is $v=0$, i.e. the strict two-jet value. Therefore $\delta v$ is simply given by the value of the event shape after gluon emission, which can be expressed in terms of the final state momenta. However at order $\alpha_s$, in the presence of a real soft emission, one needs to work out how the gluer modifies the value of $v$ relative to its non-zero pre-emission value. In order to do this we compute $\delta v$ as :
 \begin{equation}
 \label{eq:def}
 \delta  v = v(\pq,\pqbar,k_1,k_2) - v(\tildepq,\tilde{p}_{\bar{q}}, \tilde{k}_1).
 \end{equation}
where $\pq,\pqbar,k_1,k_2$ are post-emission particle four-momenta while $\tildepq,\tilde{p}_{\bar{q}}, \tilde{k}_1$ are pre-emission four-momenta i.e. before the emission of $k_2$.

 We immediately notice from Eq.~\eqref{eq:def} that determining the change in value of the event shape depends on us being able to relate the post emission values $p_q,\pqbar,k_1$ to the pre-emission ones given by $\tilde{p}_q, \tilde{p}_{\bar{q}},\tilde{k}_1$.  In order to find $\delta v$ it is necessary for us to adopt  the notion of a recoil scheme which determines how the perturbative system recoils against the non-perturbative gluon. One might consider that this introduces a further model dependence in our calculation, but it nevertheless makes it possible for us to extend the DW model to discuss the question of anomalous dimensions more concretely and gives a framework to study their universality.  We shall comment more on the choice of scheme and its potential impact on results below.

\section{Calculation for thrust mean value}
\label{sec:thrust}
We shall first carry out the standard DW style calculation.  As already commented, in order to work out the non-perturbative power correction to the mean thrust due to gluer emission, we need to evaluate the change in $1-T$ induced by gluer emission, multiply it by the gluer emission probability, and integrate over the phase space for gluer emission. While at this order, as explained before, we do not need the concrete specification of a recoil scheme, we shall already introduce our chosen scheme here, to set the scene for what follows.

The recoil mapping we adopt is a Catani-Seymour style dipole local recoil where one of the particles of the emitting dipole is designated as the ``emitter" and the other the ``spectator " \cite{Catani:1996vz}. The transverse recoil is absorbed as usual by the emitter, while the spectator takes only longitudinal recoil. The decision about which particle is designated as the emitter and which the spectator is done by considering the dipole leg closer to the gluer in angle in the {\it{event centre-of-mass frame}}.  We now proceed with calculating the leading power correction.

\subsection{Leading $1/Q$ correction}
At leading order the gluer $k$ is emitted directly from a hard $q\bar{q}$ dipole.  In the soft limit the thrust axis is along the $q\bar{q}$ axis and defines two hemispheres, where we can take the quark to be in the right hemisphere $ \mathcal{H}_R$.  In our recoil scheme, we can then parametrise the post emission momenta of particles as follows (assuming $k \in \mathcal{H}_R$ which implies that the quark is the emitter and the antiquark the spectator):
\begin{align}
\label{eq:recoil}
k &= \alpha \tildepq+\beta \tildepqbar - k_{\perp}, \\
\pq &= \alpha_q \tildepq +\beta_q \tildepqbar+k_{\perp}, \nonumber \\
\pqbar &=  \beta_{\bar{q}} \tildepqbar \nonumber,
\end{align}

where $\tildepq$ and $\tildepqbar$ are pre-emission momenta we parametrise as $\tildepq =\frac{Q}{2}(1,0,0,1),  \tildepqbar =\frac{Q}{2}(1,0,0,-1)$. Taking massless final state partons, we have that
\begin{align}
\alpha_q &=1-\alpha ,\\
\beta_q &= \frac{{\bf{k_{\perp}}}^2}{Q^2(1-\alpha)},\\
\beta_{\bar{q}} &= 1-\frac{{\bf{k_{\perp}}}^2}{Q^2 \alpha(1-\alpha)}.
\end{align}

As stated before, the thrust axis in the soft limit may be approximated by the initial $q\bar{q}$ direction. Specifically for emission into the quark hemisphere it lies along the anti-quark direction and vice-versa. Then assuming $k \in \mathcal{H}_R$ we have the following expression for the thrust, obtained by summing the modulii of the longitudinal momentum along the thrust axis :
\begin{equation}
T = \frac{\frac{Q}{2}\left(\beta_{\bar{q}}+\alpha-\beta+\alpha_q-\beta_q\right)}{Q},
\end{equation}
which gives
\begin{equation}
\tau=1-T =1-\beta_{\bar{q}} =\beta+\mathcal{O}({\bf{k_\perp}}^2/Q^2).
\end{equation}

Likewise for the case when emission is in the left hemisphere i.e. $\beta > \alpha$, we obtain $\tau=\alpha+\mathcal{O}({\bf{k_\perp}}^2/Q^2)$.
Thus in the soft approximation where we can neglect second-order terms in small quantities we have simply that  $\tau = \mathrm{min}(\alpha,\beta).$ We therefore compute the NP correction to the mean value of the thrust, from a single gluer emission as follows \footnote{To lighten notation we shall henceforth use $k_\perp^2$ to mean ${\bf{k}_\perp^2}$.}:
\begin{equation}
\langle \delta \tau \rangle^{\text{NP}}= \frac{4C_F}{\pi} \int_0^{\mu_I} \frac{d k_\perp}{k_\perp} \frac{d\alpha}{\alpha} \alpha_s(k_\perp)\ \Theta \, \left(\alpha>\frac{k_\perp}{Q} \right) \frac{k_\perp^2}{Q^2 \alpha},
\end{equation}
where the step function involving $\alpha$ restricts the gluer to $\mathcal{H}_R$ and a factor of two accounts for emission into $\mathcal{H}_L$. Here the value of $\delta v$ is given by the factor $\beta=\frac{k_\perp^2}{Q^2 \alpha}$ and we have used the standard form for the emission probability of a soft gluer in terms of $k_\perp$ and $\alpha$.
Performing the $\alpha$ integral gives:
\begin{equation}
\label{eq:dwsimp}
\langle \delta \tau \rangle^{\text{NP}} = \frac{4C_F}{Q \pi} \int_0^{\mu_I} dk_\perp \, \alpha_s(k_\perp),
\end{equation}
where we neglected terms of order $k_\perp^2/Q^2$. Eq.~\eqref{eq:dwsimp} is the well-known DW model power correction \cite{Dokshitzer_1995} which is expressed in terms of an integral of $\alpha_s(k_\perp)$ over the non-perturbative region extending up to an infrared matching scale $\mu_I$.\footnote{The full result involves also the application of the two-loop Milan factor \cite{Dokshitzer_1998,Dokshitzer_1998_2} and the subtraction to order $\alpha_s^2$ of the ill-defined perturbative contribution below the scale $\mu_I$ \cite{Dokshitzer_1995}. These details may be ignored for developing our present argument.}

\subsection{Order $\alpha_s$ correction and anomalous dimension}
In this section we shall consider gluer emission which takes place from the $q \bar{q}$ dipole dressed with an additional soft gluon with transverse momentum $k_t$ such that $Q \gg k_t \gg \Lambda$. Relative to emission off just the hard $q\bar{q}$, we now have two additional considerations. These involve i) the more complicated emitting ensemble and ii) the potentially non-trivial role of the recoil scheme. We comment on each of these in turn.

\begin{itemize}
\item Gluer emission probability:
In the limit we work in, we need to consider the emission of two soft gluons $k_1$ and $k_2$, which are strongly ordered in energy or equivalently in transverse momentum i.e. the ordered double-soft limit. In this limit there are two terms to consider which are separated by different colour factors : a $C_F^2$ abelian term amounting to independent emission from the $q\bar{q}$ dipole and a $C_F C_A$ term which can be thought of as emission of the softest gluon (gluer) from three dipoles. The $C_F C_A$ term is the same as that entering the calculation of leading non-global logarithms. A reader familiar with that calculation will be aware that in this piece there are no collinear singularities along the hard $q\bar{q}$ legs. It is this $C_F C_A$ term that shall be relevant to the anomalous dimension.

The $C_F^2$ term does not enter the anomalous dimension (or indeed NGL configurations). Here there are contributions from both the case when $k_1$ is real as well as when $k_1$ is virtual. In both cases the recoil against $k_2$ is taken by the $q\bar{q}$ dipole, which is the emitting dipole. Hence while $\delta v$ is identical in the two cases, real and virtual $k_1$ terms come with opposite signs owing to unitarity, which leads to a cancellation. This leaves only the $C_F C_A$ term to consider, where in the strongly energy ordered limit there is no contribution from a virtual $k_1$, real $k_2$ configuration \cite{DMO}. Hence in what follows we focus simply on the $C_F C_A$ part of the ordered double-soft squared matrix-element for emissions $k_1$ and $k_2$ \cite{Catani:1999ss}:
\begin{multline}
\label{eq:me}
\mathcal{M}^2_{C_F C_A}(k_1,k_2) = 2 C_F C_A \left(\frac{\alpha_s}{2\pi}\right)^2 \frac{\pq\cdot\pqbar}{(\pq\cdot k_1)(\pqbar \cdot k_1)} \left[\frac{(\pq \cdot k_1)}{(k_1 \cdot k_2) (\pq\cdot k_2)} + \right . \\  \left .+ \frac{(\pqbar\cdot k_1)}{(k_1 \cdot k_2) (\pqbar\cdot k_2)}   - \frac{\pq \cdot \pqbar }{(\pq \cdot k_2)(\pqbar \cdot k_2)}\right].
\end{multline}
The emission pattern above corresponds to the emission of a soft gluon $k_1$ from a $q\bar{q}$ dipole, followed by an emission of much softer $k_2$ from the $p_q k_1$, $p_{\bar{q}} k_1$ and $p_q p_{\bar{q}}$ dipoles. It is expressible as 

\begin{equation}
\label{mealt}
\mathcal{M}^2_{C_F C_A}(k_1,k_2) =2 C_F C_A   \left(\frac{\alpha_s}{2\pi}\right)^2 W_{q {\bar{q}}}(k_1)\sum_{\text{dipoles}(ij)}W_{ij}(k_2),
\end{equation}
with $W_{ij}(k)= \frac{(p_i.p_j)}{(p_i.k)(p_j.k)}$ being the standard dipole antenna emission pattern for an emission $k$ from a dipole with legs labeled by $i$ and $j$.

\item Recoil considerations:
In computing the change in event shape $\delta v$, we have to work out the difference between the event shape value before and after gluer emission, and the value before gluer emission is non-zero given the presence of the soft gluon $k_1$. Here unlike in the leading-order DW model, one may worry about the dependence on our choice of recoil scheme used to obtain $\delta v$. 
In our present calculation the approach we have adopted is essentially the same as that used in the appendix of Ref.~\cite{Luisoni:2020efy}, where $1/Q$ power corrections to the $C$-parameter distribution in the three-jet region were computed with a gluer emission and the concept of a recoil scheme. It was found that schemes such as the Catani-Seymour style scheme used here \cite{Catani:1996vz}, the PanLocal antenna scheme \cite{Dasgupta:2020fwr} and the PanGlobal scheme \cite{Dasgupta:2020fwr} all gave results compatible with one another. As discussed in Ref.~\cite{Luisoni:2020efy} all these schemes have the feature that longitudinal recoil is kept local within the emitting dipole. A scheme which differs in this respect is the FHP scheme \cite{Forshaw:2020wrq} which was shown to produce a different result.

Subsequent to this, studies were carried out for the three-jet limit in Ref.~\cite{Caola:2022vea} which brought substantial further insight into the role of the recoil scheme and their connection with the correct result also computed there. Crucially it was found that for recoil schemes that satisfied certain smoothness criteria, a simple soft gluon analysis was sufficient to give results that were independent of the specific scheme choice and which agreed with the correct result for linear $1/Q$ power corrections in the three-jet region. The FHP scheme did not satisfy these properties, but any of the other aforementioned schemes are evidently suitable for use within our approach here. We therefore might anticipate that while we have chosen a particular scheme for reasons of analytical simplicity, the results in schemes like the PanGlobal and PanLocal antenna schemes will yield the same results. The connection of results from these schemes with the correct results for the case of $1/Q$ corrections in the three-jet region from Refs.~\cite{Caola:2021kzt,Caola:2022vea} is also encouraging and lends justification to our approach.\footnote{We note however that strictly speaking the exact calculations performed in Refs.~\cite{Caola:2021kzt,Caola:2022vea} involve the use of a leading $n_f$ abelian theory in which there would be no anomalous dimension of the kind we compute here.}

\end{itemize}

With these remarks in place we proceed with the calculation of each of the contributing dipoles in Eq.~\eqref{eq:me}. For each dipole term we shall use our chosen recoil scheme to evaluate $\delta v$ and perform the required integrals before combining the results. We start in the next subsection by evaluating the change in thrust.

\subsection{Change in thrust due to gluer emission from dipoles}
In order to estimate the power correction induced by the emission of a gluer, we need to compute the change in the event shape due to gluer emission i.e. the quantity $\delta \tau  = \tau(\pq,\pqbar,k_1,k_2) - \tau(\tildepq,\tildepqbar, \tilde{k}_1)$, where $\tau=1-T$. 
To derive an expression for $\delta \tau$, we first identify which of the dipoles emits $k_2$, with the emission probability being given by the corresponding term of the matrix-element Eq.~\eqref{eq:me}. Then we shall use our dipole local recoil scheme to prescribe how the recoil is shared amongst the emitting legs.

Let us consider the situation where emission $k_1 \in \mathcal{H}_R$ and emission $k_2$ can be emitted into either $\mathcal{H}_L$ or $\mathcal{H}_R$, with a factor of two accounting for $k_1 \in \mathcal{H}_L$ . \footnote{In the strongly ordered limit, where we also work to first order in $k_{t_2},$ configurations where emission $k_1$ changes hemispheres due to recoil are suppressed and can be ignored.}
  In the limit where the emissions are soft and strongly-ordered in energy, $Q \gg \omega_1\gg \omega_2$, the thrust axis, set by the 3 hardest emissions, lies along the initial $q \bar{q}$ direction, specifically along $\pqbar$. Then given that the thrust is given by the (normalised) modulus of the longitudinal momentum along this axis we have
\begin{align}
T(\pq,\pqbar,k_1,k_2) &= \frac{1}{Q} \left(|p_{q,L}|+|k_{1,L}|+|k_{2,L}|+|p_{\bar{q},L}| \right), \\
T(\tildepq,\tildepqbar,\tilde{k}_1) &= \frac{1}{Q} \left(|\tilde{p}_{q,L}|+|\tilde{k}_{1,L}|+|\tilde{p}_{\bar{q},L}| \right).
\end{align}

Particles in $\mathcal{H}_R$ have a negative longitudinal component of momentum along the thrust axis (anti-quark direction) while those in $\mathcal{H}_L$ have positive components.
Accordingly for any particle $p$ we may replace $|p_L|$ by $p_L$ if the particle is in $\mathcal{H}_L$ or by $- p_L$ if it is in $\mathcal{H}_R$. Then we have 
\begin{equation}
\label{eq:delta}
\delta T = T(\pq,\pqbar,k_1,k_2)-T(\tildepq,\tildepqbar,\tilde{k}_1) = \frac{1}{Q} \left( \tilde{p}_{q,L}-p_{q,L}+\tilde{k}_{1,L}-k_{1,L}-\tilde{p}_{{\bar{q}},L}+p_{{\bar{q}},L}+|k_{2,L} | \right).
\end{equation}
where we expressed the result in terms of the components of particle momenta along the thrust axis and left the dependence on the modulus of $k_{2,L}$ since this is either  $k_{2,L}$ or $-k_{2,L}$ depending on whether $k_2 \in \mathcal{H}_L$ or $k_2 \in \mathcal{H}_R$. From momentum conservation we also have:

\begin{equation}
\label{eq:momcorr}
p_{q,L}+k_{1,L}+k_{2,L}+p_{{\bar{q}},L} = \tilde{p}_{q,L}+\tilde{k}_{1,L}+ \tilde{p}_{{\bar{q}},L} = 0.
\end{equation}

We now have three scenarios each of which yields a different expression for $\delta T$:

\begin{itemize} 
\item a) Emission from the $\pq k_1$ dipole:\\
Here $p_{\bar{q}}$ is not involved in the emission at all and we have that $p_{{\bar{q}},L}= \tilde{p}_{{\bar{q}},L}$. Putting together eqs.~\eqref{eq:delta} and \eqref{eq:momcorr} we get
\begin{equation}
\label{eq:p1k1}
\delta T = \frac{1}{Q} \left(k_{2,L}+|k_{2,L}| \right) = \frac{2}{Q} k_{2,L} \Theta_{k_2 \in \mathcal{H}_L}.
\end{equation}

\item b) Emission from the $\pqbar k_1$ dipole:\\
Here $p_q$ is not involved in the emission at all and we have that $p_{q,L}= \tilde{p}_{q,L}$. Putting together eqs.~\eqref{eq:delta} and \eqref{eq:momcorr} we can write the result in the following equivalent ways:

\begin{align}
\label{eq:p2k1-k1}
\delta T &= \frac{2}{Q} \left( \Theta_{k_2 \in \mathcal{H}_L} \left( p_{\bar{q},L}- \tilde{p}_{\bar{q},L} +k_{2,L} \right)+  \Theta_{k_2 \in \mathcal{H}_R} \left( p_{\bar{q},L}- \tilde{p}_{\bar{q},L}  \right)\right) \\
 &= \frac{2}{Q} \left( \Theta_{k_2 \in \mathcal{H}_L} \left(\tilde{k}_{1,L}-k_{1,L} \right)+  \Theta_{k_2 \in \mathcal{H}_R} \left(\tilde{k}_{1,L}-k_{1,L} -k_{2,L}\right)\right) \label{eq:p2k1-p2}.
\end{align}

We shall use both of the forms above in the calculation to follow, depending on which particle is designated as the emitter and which one the spectator in the dipole recoil scheme we adopt.

\item c) Emission from the $\pq \pqbar$ dipole: \\
Here $k_1$ is not involved in the emission at all and we have that $k_{1,L}= \tilde{k}_{1,L}$. Putting together eqs.~\eqref{eq:delta} and \eqref{eq:momcorr} we can write the result as
\begin{equation}
 \frac{2}{Q} \left( \Theta_{k_2 \in \mathcal{H}_L} \left( \tilde{p}_{q,L}- p_{q,L} \right)+  \Theta_{k_2 \in \mathcal{H}_R} \left( p_{{\bar{q}},L}-  \tilde{p}_{{\bar{q}},L}  \right)\right). \\
\end{equation}
\end{itemize}

\subsubsection{Expressions for $\delta \tau$ from dipole recoil scheme}
Having written the general form for the change in thrust in terms of the components of longitudinal momenta of the emitting particles and the emission $k_2$, we now derive explicit expressions in terms of the phase-space variables, using our recoil scheme. Given that this is a dipole style recoil scheme, where one particle is designated as emitter and absorbs the transverse recoil while the other particle is the spectator which only takes longitudinal recoil as prescribed in eq.~\eqref{eq:recoil}, it proves advantageous to use the form for the $\delta T$ written in terms of the momentum of the spectator and the emission $k_2$. We also remind the reader that in deriving the expressions below we are working in an approximation designed to capture the linear power correction in $k_2$ and to leading logarithmic order in $k_1$ i.e. neglecting all higher order terms contributing beyond these limits.

Let us start with the simplest case, case a), which is emission from the $p_q k_1$ dipole.
The result is given in \eqref{eq:p1k1} and the quantity $k_{2,L}$, i.e. the longitudinal momentum of $k_2$, is easily expressed in terms of the energy fraction $x_2$ and the angle $\theta_{2q}$ between emission $k_2$ and the quark. Accounting for a change of sign stemming from the fact that we want the result for $\delta \tau$ rather than $\delta T$, we obtain
\begin{equation}
\label{eq:p1k1xt}
\delta \tau^{p_q k_1} = x_2  \cos \theta_{2q}  \,  \Theta \left(\theta_{2q}> \frac{\pi}{2} \right),
\end{equation}
where the step function takes care of emission $k_2$ belonging in $\mathcal{H}_L$ and $x_2$ is the usual energy fraction $x_2 = 2 E_2/Q$, with $E_2$ the energy of $k_2$.

Next we consider emission from the $\pqbar k_1$ dipole. Here we use either expression \eqref{eq:p2k1-k1} when $k_1$ is the emitter or  \eqref{eq:p2k1-p2} when $\pqbar$ is the emitter, supplying an additional minus sign for $\delta \tau$ relative to the case of $\delta T$. When $k_1$ is the emitter we therefore need to evaluate $\tilde{p}_{\bar{q},L}-p_{\bar{q},L}$. This is simple since $\pqbar$ is the spectator so its recoil against $k_2$ is given by $\tildepqbar \to \pqbar= \beta_{{\bar{q}}} \tildepqbar$. 
\begin{equation}
\delta \tau ^{\pqbar k_1, k_1 \, \text{emitter}} =  \frac{2}{Q} \left( \Theta_{k_2 \in \mathcal{H}_L} \left( \tilde{p}_{{\bar{q}},L}(1-\beta_{\pqbar}) -k_{2,L} \right)+  \Theta_{k_2 \in \mathcal{H}_R} \left( \tilde{p}_{{\bar{q}},L}(1-\beta_{\pqbar})  \right)\right).
\end{equation}
The quantity $\beta_{\pqbar}$ is given by momentum conservation and we have
\begin{equation}
1-\beta_{\pqbar} = \frac{k_1 \cdot k_2}{\tildepqbar \cdot \tilde{k}_1},
\end{equation}
i.e. the ratio of the invariant mass of the emission and the emitter (post emission) to the invariant mass of the emitting dipole (pre-emission).

In evaluating $\beta_{\pqbar}$, working to leading order in the softness of emission $k_2$,  as needed for the linear power correction, we may replace $\tildepqbar$, $\tilde{k}_1$ by $\pqbar$ and $k_1$ respectively. Hence in terms of our integration variables, i.e. the energies and angles, we obtain 
\begin{equation}
1-\beta_{\pqbar} = x_2 \left( \frac{1-\cos\theta_{12}}{1+\cos\theta_{1q}} \right),
\end{equation}
where $\theta_{12}$ is the angle between $k_1$ and $k_2$ and $\theta_{1q}$ that between $k_1$ and the quark and we have taken $\theta_{2\bar{q}}=\pi-\theta_{2q}$.
\footnote{We can take the antiquark to be back to back wrt the quark direction, as any transverse recoil against $k_1$ will only impact terms beyond the soft leading logarithm we wish to isolate.}
Hence we get 
\begin{multline}
\label{eq:k1}
\delta \tau^{\pqbar k_1,\, k_1 \text{emitter}} = x_2 \left(\frac{1-\cos\theta_{12}}{1+\cos\theta_{1q}}\right) \Theta \left(\theta_{2q} < \frac{\pi}{2} \right)+ \left [x_2 \left(\frac{1-\cos\theta_{12}}{1+\cos\theta_{1q}}\right) \right.+\\
\left.+x_2 \cos \theta_{2q} \right  ] \Theta \left(\theta_{2q} > \frac{\pi}{2} \right).
\end{multline}

Similarly when $\pqbar$ is the emitter we need the longitudinal recoil of the spectator $k_1$, as given by $k_{1,L}-\tilde{k}_{1,L} = (\beta_{k_1}-1) \tilde{k}_{1,L}$

\begin{equation}
1-\beta_{k_1}= \frac{\pqbar \cdot k_2}{\pqbar \cdot k_1} = \frac{x_2}{x_1} \frac{(1+\cos\theta_{2q})}{(1+\cos \theta_{1q})},
\end{equation}
which using Eq.~\eqref{eq:p2k1-p2} finally gives us 

\begin{multline}
\label{eq:pqbar}
\delta \tau^{\pqbar k_1,\, \pqbar \text{emitter}} =  \Theta \left(\theta_{2q} < \frac{\pi}{2} \right) \left [x_2\frac{(1+\cos\theta_{2q})}{(1+\cos \theta_{1q})} \cos\theta_{1q}-x_2 \cos \theta_{2q}\right ]+\\
+\left[x_2\frac{(1+\cos\theta_{2q})}{(1+\cos \theta_{1q})} \cos\theta_{1q}\ \right] \Theta \left(\theta_{2q} > \frac{\pi}{2} \right).
\end{multline}

For the case of the $\pq,\pqbar$ dipole we obtain 

\begin{equation}
\label{eq:qqbar}
\delta \tau^{\pq \pqbar} = \Theta  \left(\theta_{2q} < \frac{\pi}{2} \right) \frac{x_2}{2} \left(1-\cos\theta_{2q}\right)+\Theta  \left(\theta_{2q} > \frac{\pi}{2} \right) \frac{x_2}{2} \left(1+\cos\theta_{2q}\right),
\end{equation}
in writing which we have exploited the fact that, given our recoil prescription, the emission $k_2$ has $\pq$ as emitter if emitted into $\mathcal{H}_R$ and $\pqbar$ as emitter if emitted into 
$\mathcal{H}_L$.

Note that for the $\bar{p}_q k_1$ dipole case we need to choose an expression depending on the choice of which leg acts as emitter and which leg as spectator. In our chosen scheme we remind the reader that this is given by determining the angle of each leg with $k_2$, where we take the leg closer in angle to be the emitter. We shall apply this angular constraint in evaluating the integrals below.  We can now proceed to evaluate the integrals which give the non-perturbative correction for the change in $1-T=\tau$ associated to gluer emission off the three-particle perturbative system. 

\subsection{Evaluation of integrals}

We start by parametrising the momenta of emissions $k_1$ and $k_2$. In evaluating the squared matrix-element we can just use the soft limit (neglecting recoil of hard partons $\pq$ and $\pqbar$ against soft emissions, as this recoil will lead to terms that  are higher-order in small quantities, relative to the leading term we compute). Then taking $p_{q/\bar{q}}=Q/2(1,0,0,\pm1)$ with the quark along the positive $z$ direction, we can write
\begin{align}
k_1 &= \frac{Q}{2} x_1  \left(1,0,\sin\theta_{1q},\cos\theta_{1q} \right) ,\\
k_2 &= \frac{Q}{2} x_2 \left(1,\sin\theta_{2q} \sin \phi,\sin\theta_{2q} \cos \phi,\cos\theta_{2q} \right), 
\end{align}
where the polar angle $\theta_{iq}$ refers to the angle of emission $i$ wrt the quark direction. 

We now consider the integrals from various dipole contributions for the emission of the gluer $k_2$. The general evaluation for the gluer induced contribution to the mean value of $1-T$ then takes the form (recall the ordering $x_2 \ll x_1$):
\begin{multline}
\label{eq:dipoledef}
 \langle  \delta \tau \rangle^{\text{NP},1} = C_F C_A \frac{\alpha_s}{2\pi} \sum_{\text{dipoles}\, (ij)} \frac{Q^4}{16} \int_0^1  x_1 dx_1  x_2 dx_2\, \theta(x_1>x_2) \int_{-1}^{1} d\cos{\theta_{1q}}  d\cos{\theta_{2q}}\int_0^{2\pi} \frac{d\phi}{2\pi} \times \\ \int_0^{\mu_I} d\kappa_{T} \delta \left(\kappa_{T,ij}-\kappa_T\right)
\frac{\alpha_s(\kappa_{T,ij})}{2\pi} \, W_{q\bar{q}}(k_1) W_{ij}(k_2) \, \delta \tau^{p_i p_j} (x_1,x_2,\theta_{1q},\theta_{2q},\phi),
\end{multline}
where the dipole transverse momentum for an emission $k$ is given by:
\begin{equation}
 \kappa_{T,ij}=\sqrt{2\frac{(p_i\cdot k)(p_j \cdot k)}{(p_i \cdot p_j)}},
 \end{equation}
and where we have used the notation $ \langle  \delta \tau \rangle^{\text{NP},1}$ to denote the leading-order in $\alpha_s$ log enhanced correction to the pure $1/Q$ power behaved term. 

The above Eq.~\eqref{eq:dipoledef} accounts for a number of key considerations. It represents an evaluation, to our required accuracy, of the non-perturbative contribution to the shape mean value as given in Eq.~\eqref{eq:schem}, using the squared matrix-element of Eqs.~\eqref{eq:me}, \eqref{mealt}. The result takes the form of a sum over dipole contributions, with a $\delta \tau^{p_ip_j}$ factor for each dipole, as computed in the previous sections. As for the standard DW calculation, we carry out the phase-space integrals at fixed dipole transverse momentum, 
$\kappa_{T,ij} =\kappa_T$ for the gluer $k_2$. The final result involves integrating over $\kappa_T$ in the non-perturbative region below $\mu_I$. Finally we have used a fixed-coupling approximation for the emission of the perturbative gluon $k_1$. Beyond the strict fixed-order approach we use here, one needs to consider running of the coupling for emission $k_1$, over the transverse momentum range between $\kappa_T$ and $Q$. A resummed calculation would necessarily account for this running.

In order to evaluate Eq.~\eqref{eq:dipoledef} we can use the expressions for the gluer induced change in thrust $\delta \tau^{p_i p_j}$, which we reported in Eqs.~\eqref{eq:p1k1xt}--\eqref{eq:qqbar}. In general we can see that determining the explicit forms for each parent dipole, within our choice of recoil scheme, requires us to specify the emitter and spectator. 

Therefore we can write
\begin{equation}
\label{eq:dipolethrust}
\delta \tau^{p_i p_j} (x_1,x_2,\theta_{1q},\theta_{2q},\phi) = \delta \tau^{p_i p_j,\,p_i \,\, \text{emitter}} \theta \left( \theta_{2i}< \theta_{2j} \right)+ \delta\tau^{p_i p_j,\,p_j \,\, \text{emitter}} \theta \left( \theta_{2i}> \theta_{2j} \right)
\end{equation}
where we have designated the emitter and spectator based simply on closeness in angle. We now proceed with evaluating each individual contribution.

\subsubsection{Emission from the $p_q k_1$ dipole} 
Here we consider the first of three dipole emission terms in the squared matrix-element i.e. the contribution involving the first term in square brackets on the RHS of \eqref{eq:me}.  For this term we obtain
 \begin{multline}
 \langle  \delta \tau \rangle^{\text{NP},1,p_q k_1} =8 C_F C_A \frac{\alpha_s}{2\pi} \int_0^1 \frac{dx_1}{x_1}\frac{dx_2}{x_2} \theta(x_1>x_2) \int_{0}^{1} d\cos{\theta_{1q}} \int_{-1}^0 d\cos{\theta_{2q}}\int_0^{2\pi} \frac{d\phi}{2\pi} d\kappa_{T}  \\\frac{\alpha_s(\kappa_T)}{2\pi} \times \frac{x_2 \cos \theta_{2q}}{\left(1-\cos\theta_{12} \right) (1+\cos\theta_{1q})(1-\cos\theta_{2q})} \delta \left(\kappa_T - \frac{Q}{2} x_2 \sqrt{\frac{2(1-\cos\theta_{12})(1-\cos\theta_{2q})}{1-\cos\theta_{1q}}} \right).  \\
\end{multline}

On the second line the denominator in the angular factor comes from the relevant contribution to the squared matrix-element $W_{q k_1}(k_2)$. It has collinear enhancements when $k_2$ is collinear to either of the dipole legs i.e. in the limits $\theta_{12} \to 0$ and $\theta_{2q} \to 0$. The factor of $x_2 \cos \theta_{2q}$ in the numerator comes directly from the change in thrust reported in Eq.~\eqref{eq:p1k1xt}. Note that in this particular case one does not need to separate the emitter and spectator contributions and the result can be entirely expressed in terms of the longitudinal momentum of the emission $k_2$.  The angular integration range is restricted by the fact we have placed $k_1$ in $\mathcal{H}_R$, while $k_2$ is emitted with $\theta_{2q} > \pi/2$ due to Eq.~\eqref{eq:p1k1xt}. We have also included a factor of two to account for the case when $k_1$ would have been in $\mathcal{H}_L$.

We can use the delta function defining 
$\kappa$ to perform the trivial integral over $x_2$. The integral over $x_1$ is also straightforward and, due to the ordering condition $1 \gg x_1 \gg x_2 \sim \kappa_T/Q$, it simply produces a logarithm of the form $\ln Q/\kappa_T$. Subleading terms i.e. those not enhanced by a logarithm of $Q/\kappa_T$ are dropped here.

Thus we obtain

\begin{equation}
\label{eq:p1k1int}
 \langle  \delta \tau \rangle^{\text{NP},1,p_q k_1} =  8 C_F C_A \frac{\alpha_s}{2\pi} \int_0^{\mu_I} d\kappa_T \frac{\alpha_s(\kappa_T)}{2\pi} \ln \frac{Q}{\kappa_T} \times I^{p_q k_1},
\end{equation}

where
\begin{equation}
\label{eq:pqk1int}
I^{p_q k_1} =  \int_{0}^{1} d\cos{\theta_{1q}} \int_{-1}^0 d\cos{\theta_{2q}}\int_0^{2\pi} \frac{d\phi}{2\pi}  \frac{\sqrt{2} \cos \theta_{2q}\sqrt{1-\cos\theta_{1q}}}{\left(1-\cos\theta_{12} \right)^{3/2} (1+\cos\theta_{1q})(1-\cos\theta_{2q})^{3/2}}.
\end{equation}

\subsubsection{Emission from the $\pqbar k_1$ dipole}
Following the same steps as for the $\pq k_1$ dipole we can write 
\begin{equation}
\label{eq:pqbark1}
 \langle  \delta \tau \rangle^{\text{NP},1,p_{\bar{q}} k_1} =  8 C_F C_A \frac{\alpha_s}{2\pi} \int_0^{\mu_I} d\kappa_T \frac{\alpha_s(\kappa_T)}{2\pi} \ln \frac{Q}{\kappa_T} \times I^{p_{\bar{q}} k_1},
\end{equation}

where 

\begin{equation}
I^{p_{\bar{q}} k_1} =  \int_{0}^{1} d\cos{\theta_{1q}} \int_{-1}^{1} d\cos{\theta_{2q}}\int_0^{2\pi} \frac{d\phi}{2\pi}  \delta \tau^{\pqbar k_1}  \frac{\sqrt{2}\sqrt{1+\cos\theta_{1q}}}{(1+\cos\theta_{2q})^{3/2} (1-\cos\theta_{1q}) \left(1-\cos\theta_{12} \right)^{3/2}}.
\end{equation}

\subsubsection{Emission from the $p_q,\pqbar$ dipole} 
Here we obtain 
\begin{equation}
 \langle  \delta \tau \rangle^{\text{NP},1,p_q p_{\bar{q}}}  =  8 C_F C_A \frac{\alpha_s}{2\pi} \int_0^{\mu_I} d\kappa_T \frac{\alpha_s(\kappa_T)}{2\pi} \ln \frac{Q}{\kappa_T} \times I^{p_q p_{\bar{q}}},
\end{equation}
where 
\begin{equation}
I^{p_{\bar{q}} p_q} =  -\int_{0}^{1} d\cos{\theta_{1q}} \int_{-1}^{1} d\cos{\theta_{2q}}\int_0^{2\pi} \frac{d\phi}{2\pi}  \delta \tau^{p_q \pqbar} \frac{4}{(1-\cos^2 \theta_{1q})(1-\cos^2\theta_{2q})^{3/2}},
\end{equation}
where we note the negative sign in the definition of $I^{p_{\bar{q}} p_q} $ which results from the negative sign associated to the $q\bar{q}$ dipole term in the $C_F C_A$ piece of the ordered double soft matrix-element \eqref{eq:me}.
\subsubsection{Numerical results}
Here we perform the numerical calculations for the dipole integrals defined above. Let us start with the contribution from the $p_q k_1$ dipole where the integral to evaluate is given in eq.~\eqref{eq:pqk1int}.
We can express $1-\cos \theta_{12}$ in terms of $\cos \theta_{1q}, \cos \theta_{2q}$ and $\phi$ and perform the integrals numerically to obtain
\begin{equation}
\label{eq:Ip1k1}
I^{p_q k_1} = -0.42788830.
\end{equation}

Now we turn to the contributions from the $\pqbar k_1$ and $p_q \pqbar$ dipole emission terms, Prior to evaluating these integrals we recall that when all terms of the matrix-element are combined, we are able to integrate freely over the directions of gluons $k_1$ and $k_2$ without obtaining collinear divergences. The correlated emission piece of the double-soft matrix-element that we are focussing on here is free from collinear singularities along $p_q$ and $\pqbar$, when all terms are combined. The only genuine collinear singularity present is for the case when $k_1$ becomes collinear to $k_2$, but in this limit the collinear safety of the thrust variable ensures that we have a finite result. However in evaluating the individual contributions separated as above, one encounters a collinear divergence for the situation when emission $k_1$ goes collinear to $p_q$, i.e. when $\theta_{1q}$ vanishes.\footnote{The divergences are in any case cut-off by the fact that the transverse momentum of emission $k_1$ should be larger than that of $k_2$ which gives a cut-off on $\theta_{1q}$. If one retains such a cut-off the collinear divergence will become a collinear logarithmic enhancement. Since ultimately these collinear enhancements cancel between different dipole contributions, we can set the cut-off to zero from the start.} This divergence obviously cancels in the sum over dipole contributions from the $p_q \pqbar$ and $\pqbar k_1$ dipole cases described above. Hence it is useful to combine these two contributions from the outset. Doing so will let us write the result as a sum of individually finite integrals.

 In order to efficiently combine the $p_q \pqbar$ and $\pqbar k_1$ dipole cases, it proves helpful to consider the cases when gluon $k_2$ is emitted into $\mathcal{H}_R$ or  $\mathcal{H}_L$. The situation for the $p_q \pqbar$ dipole is simple : if the gluon is emitted in  $\mathcal{H}_L$ the emitter is $\pqbar$ otherwise it is $p_q$. For emission from the  $\pqbar k_1$ dipole the situation is more complicated as one has to find the angle between $k_2$ and the dipole legs to determine the emitter. This leads us to introduce a simplified version of the calculation where if $k_2$ is emitted into $\mathcal{H}_L$ we shall designate $\pqbar$ as the emitter as is the case for the $p_q \pqbar$ dipole, otherwise we designate $k_1$ as the emitter. The simplified version of the $\pqbar k_1$ contribution combines together with the $p_q \pqbar$ contribution to give a collinear finite integral, which is easily evaluated numerically. This leaves us to evaluate a correction term which accounts for the difference between the simplified and correct contributions from  $\pqbar k_1$, which is also collinear finite and easily evaluated numerically.
 
 The first step therefore is to split 
 \begin{equation}
 \label{eq:split}
\delta \tau^{\pqbar k_1} = \delta \tau^{\pqbar k_1, \text{simp.}}+ \delta \tau^{\pqbar k_1, \text{corr.}},
 \end{equation}
 where as explained above
 \begin{equation}
 \label{eq:split2}
 \delta\tau^{\pqbar k_1,\text{simp.}} = \delta \tau^{\pqbar k_1, \pqbar \, \text{emitter}} \theta\left(\theta_{2q}>\frac{\pi}{2}\right) +  \delta \tau^{\pqbar k_1, k_1 \, \text{emitter}} \theta \left(\theta_{2q}<\frac{\pi}{2}\right),
 \end{equation}
 where we have designated $\pqbar$ as emitter for $k_2 \in \mathcal{H}_L$ and $k_1$ as emitter otherwise. From Eqs.~\eqref{eq:k1},\eqref{eq:pqbar} we obtain 
 \begin{multline}
\delta\tau^{\pqbar k_1, \text{simp.}} =  \left[x_2\frac{(1+\cos\theta_{2q})}{(1+\cos \theta_{1q})} \cos\theta_{1q}\ \right] \Theta \left(\theta_{2q} > \frac{\pi}{2} \right)+\\+\left [x_2 \left(\frac{1-\cos\theta_{12}}{1+\cos\theta_{1q}}\right) \right  ] \Theta \left(\theta_{2q} < \frac{\pi}{2} \right).
\end{multline}
 We shall now combine the contributions from the $p_q \pqbar$ dipole and the simplified $\pqbar k_1$ contribution above. For $k_2 \in \mathcal{H}_L$ we have :
 \begin{multline}
 I^{\pqbar k_1, \text{simp.},k_2 \in \mathcal{H}_L}+I ^{p_q \pqbar, k_2 \in \mathcal{H}_L} = \int_{0}^{1} d\cos{\theta_{1q}} \int_{-1}^{0} d\cos{\theta_{2q}}\int_0^{2\pi} \frac{d\phi}{2\pi} \times \\
 \left[\frac{\sqrt{2}}{\sqrt{(1+\cos\theta_{1q})(1+\cos\theta_{2q})}}  \frac{\cos\theta_{1q}}{1-\cos\theta_{1q}} \frac{1}{(1-\cos\theta_{12})^{3/2}}+\right. \\ \left. -\frac{2}{(1-\cos^2\theta_{1q}) \sqrt{1+\cos\theta_{2q}} (1-\cos\theta_{2q})^{3/2}}\right],
 \end{multline}
 which we can evaluate numerically:
 \begin{equation}
 \label{eq:Ipqbark1a}
 I^{\pqbar k_1, \text{simp.},k_2 \in \mathcal{H}_L}+I ^{p_q \pqbar,k_2 \in \mathcal{H}_L} = -0.2866603285.
\end{equation} 

For the situation when $k_2$ is emitted in $\mathcal{H}_R$ we have to evaluate
\begin{multline}
I^{\pqbar k_1, \text{simp.},k_2 \in \mathcal{H}_R}+I ^{p_q \pqbar, k_2 \in \mathcal{H}_R} = \int_{0}^{1} d\cos{\theta_{1q}} \int_{0}^{1} d\cos{\theta_{2q}}\int_0^{2\pi} \frac{d\phi}{2\pi} \frac{1}{(1-\cos\theta_{1q})(1+\cos\theta_{2q})^{3/2}} \times\\ \left[ \frac{\sqrt{2}}{ \sqrt{1+\cos\theta_{1q}(1-\cos\theta_{12})}}  -\frac{2}{1+\cos\theta_{1q}} \frac{1}{\sqrt{1-\cos\theta_{2q}}}\right],\\
  \end{multline}
which gives 
\begin{equation}
\label{eq:Ipqbark1b}
 I^{\pqbar k_1, \text{simp.},k_2 \in \mathcal{H}_R}+I ^{p_q \pqbar, k_2 \in \mathcal{H}_R} = -0.2870299785.
\end{equation}

Finally we evaluate the contribution from the term that corrects for our simplified treatment of the emitter and spectator in the $\pqbar k_1$ dipole emission. Here when emission $k_2$ is emitted in the right hemisphere, we had designated $k_1$ as emitter but this is actually only correct if the emission $k_2$ is closer in angle to $k_1$. In the region when it is closer in angle to $\pqbar$, the correct emitter is $\pqbar$ and hence we need a correction term which accounts for the difference between the true and simplified situation. Similarly when $k_2$ is emitted in the left hemisphere, when closer in angle to $k_1$ the true emitter is $k_1$ rather than $\pqbar$ as assumed in the simplified case. 

Hence still in the context of Eq.~\eqref{eq:pqbark1} we identify two correction terms according to whether $k_2 \in \mathcal{H}_L$ or $k_2 \in \mathcal{H}_R$. These terms arise from the correct form of $\delta \tau$ versus its simplified form which yields, using Eqs.~\eqref{eq:split}, \eqref{eq:split2}:

\begin{multline} 
\delta \tau^{\pqbar k_1, \text{corr.}}  =   \theta \left( \theta_{2q}<\frac{\pi}{2} \right)  \theta(\theta_{2\bar{q} }<\theta_{12}) \left(\delta \tau^{\pqbar k_1, \pqbar\text{emitter}}-\delta \tau^{\pqbar k_1, \, k_1 \text{emitter}}\right) +\\ 
+    \theta \left( \theta_{2q}>\frac{\pi}{2} \right)  \theta(\theta_{2\bar{q}}>\theta_{12})  \left(\delta \tau^{\pqbar k_1, k_1 \text{emitter}}-\delta \tau^{\pqbar k_1, \, \pqbar \text{emitter}}\right).
\end{multline}

The explicit expressions for the various terms on the RHS of the above equation are given in Eqs.~\eqref{eq:k1}, \eqref{eq:pqbar}. Inserting the form of $\delta \tau^{\pqbar k_1, \text{corr.}}$ in Eq.~\eqref{eq:pqbark1} for emission from the $\pqbar k_1$ dipole we obtain the following additional contributions due to the correction terms (for the cases when $k_2$ is in $\mathcal{H}_R$ or in $\mathcal{H}_L$ respectively) :
\begin{multline}
 I^{\pqbar k_1, \text{corr.},k_2 \in \mathcal{H}_R} =   \int_{0}^{1} d\cos{\theta_{1q}} \int_{0}^{1} d\cos{\theta_{2q}}\int_0^{2\pi} \frac{d\phi}{2\pi}  \frac{\sqrt{2}\sqrt{1+\cos\theta_{1q}}}{(1+\cos\theta_{2q})^{3/2} (1-\cos\theta_{1q}) \left(1-\cos\theta_{12} \right)^{3/2}} \times \\ \times \left[ \frac{1+\cos\theta_{2q}}{1+\cos\theta_{1q}} \cos\theta_{1q}-\left(\frac{1-\cos\theta_{12}}{1+\cos\theta_{1q}} +\cos\theta_{2q}\right) \right]  \theta(\theta_{2\bar{q}}< \theta_{12}),
\end{multline}
where $\theta_{2\bar{q}}=\pi-\theta_{2q}$, and 

\begin{multline}
 I^{\pqbar k_1, \text{corr.},k_2 \in \mathcal{H}_L} =   \int_{0}^{1} d\cos{\theta_{1q}} \int_{-1}^{0} d\cos{\theta_{2q}}\int_0^{2\pi} \frac{d\phi}{2\pi}  \frac{\sqrt{2}\sqrt{1+\cos\theta_{1q}}}{(1+\cos\theta_{2q})^{3/2} (1-\cos\theta_{1q}) \left(1-\cos\theta_{12} \right)^{3/2}} \times \\ \times  \left[\frac{1-\cos\theta_{12}}{1+\cos\theta_{1q}} -\left(\frac{1+\cos\theta_{2q}}{1+\cos\theta_{1q}} \cos\theta_{1q}-\cos\theta_{2q} \right) \right] \theta(\theta_{2\bar{q}}> \theta_{12}).
\end{multline}

The integrals can be evaluated numerically and we quote below the overall result for the correction terms:
\begin{equation}
\label{eq:Icorr} 
 I^{\pqbar k_1, \text{corr.}}= I^{\pqbar k_1, \text{corr.},k_2 \in \mathcal{H}_R}+ I^{\pqbar k_1, \text{corr.},k_2 \in \mathcal{H}_L}= -1.453482423.
\end{equation}

\subsection{Final result for thrust}
We are now in a position to combine all terms and obtain our final result for the thrust. Combining the contributions reported in Eqs.~\eqref{eq:Ip1k1}, \eqref{eq:Ipqbark1a}, \eqref{eq:Ipqbark1b}, \eqref{eq:Icorr} gives
\begin{equation}
\label{eq:thrustres}
 \langle  \delta \tau \rangle^{\text{NP},1} =   -C_F C_A  \frac{\alpha_s}{2\pi}  \frac{1}{Q} \int_0^{\mu_I} d\kappa_T \frac{\alpha_s(\kappa_T)}{2\pi} \ln \frac{Q}{\kappa_T} \times 19.64048824,
\end{equation}
where the numerical coefficient quoted above arises from summing the results in  Eqs.~\eqref{eq:Ip1k1}, \eqref{eq:Ipqbark1a}, \eqref{eq:Ipqbark1b}, \eqref{eq:Icorr} and multiplying by the factor of 8 present in the prefactor for all dipole terms (see e.g. Eq.~\eqref{eq:p1k1int} for the $p_q k_1$ dipole case).

The above result shows that one obtains a $1/Q$ correction with a perturbatively calculable coefficient and a logarithmic enhancement involving the ratio of the hard scale $Q$ and the gluer transverse momentum $\kappa_T$. The scale $\kappa_T$ is integrated over up to the infrared scale $\mu_I$ which is of the order of the QCD scale $\Lambda$, using the standard DW universal non-perturbative coupling. The accompanying coefficient resulting from the angular dipole integrals is sizeable. In the following section we perform similar calculations for the case of the $C$-parameter.

\section{Calculation for $\langle C \rangle$}
\label{sec:c}
We start with some basic expressions for the $C$-parameter in the limit of multiple soft emissions. For an initial $q\bar{q}$ pair and any number of additional soft emissions we can express the result as 
\begin{equation}
C = \sum_i 6 \frac{\alpha_i \beta_i}{\alpha_i+\beta_i} = \frac{3}{Q} \sum_i \frac{k_{\perp,i}^2}{E_i},
\end{equation}
where $\alpha_i, \beta_i$ are the usual Sudakov components along light-like vectors parallel to the thrust axis in the soft limit, while $k_{\perp,i}$ and $E_i$ are respectively the  magnitude of the transverse momentum (wrt the thrust axis) and energy of emission $i$. 

\subsection{Leading-order result}
At leading order, similar to the thrust case, evaluating the change in $C$ due to gluer emission does not depend on a recoil scheme given that  $C=0$ at the Born level. Hence we obtain
\begin{align}
\label{eq:loc}
\langle \delta C \rangle^{\text{NP}} &= \frac{4 C_F}{\pi} \int_0^{\mu_I} \frac{d k_{\perp}}{k_\perp} \frac{d\alpha}{\alpha} \alpha_s(k_\perp)\ \Theta \, \left(\alpha>\frac{k_\perp}{Q} \right) \times 6 \alpha \frac{k_\perp^2/Q^2}{(k_\perp^2/Q^2+ \alpha^2)} \\ 
                                                        & \approx \frac{4 C_F}{\pi Q} \int_0^{\mu_I} dk_\perp \alpha_s(k_\perp) \times\frac{3\pi}{2}  = \frac{3\pi}{2} \langle \delta \tau \rangle^{\text{NP}},
\end{align}
where after performing the $\alpha$ integral and neglecting higher-order power corrections varying as $k_\perp^2/Q^2$, we arrived at the known result for the  $1/Q$ power correction to the $C$-parameter, which has a coefficient  $\frac{3\pi}{2}$ relative to the DW result for the thrust, Eq.~\eqref{eq:dwsimp}.

\subsection{Two gluon case : $\delta C$ evaluation}
We now move on to studying the anomalous dimension for the $1/Q$ correction. Similar to the thrust case we need to work out the change $\delta C$ induced by emission of a gluer $k_2$, given a soft perturbative emission $k_1$. Here, as before, we will need to use our recoil scheme to work out the quantity:
\begin{equation}
\delta C = C(p_q, \pqbar,k_1,k_2) -C(\tilde{p}_q,\tilde{p}_{\bar{q}},\tilde k_1),
\end{equation}
which gives the change in $C$ due to gluer emission. To evaluate this most straightforwardly we shall find it helpful to express the $C$-parameter in terms of the energies $E_i$ and longitudinal momenta $k_{L,i}$ of soft emissions:
\begin{equation}
C =  \frac{3}{Q} \sum_i \frac{k_{\perp,i}^2}{E_i} = \frac{3}{Q} \sum_i \left(E_i -\frac{k_{L,i}^2}{E_i} \right).
\end{equation}

For the evaluation of $\delta C$ we need to take the difference between the four-parton and three-parton configurations i.e. the values of $C$ after and before emisssion of $k_2$ which gives us: 
\begin{equation}
\label{eq:delc0}
\delta C =  \frac{3}{Q} \left [E_2- \frac{k_{L,2}^2}{E_2} +\delta E_1 -  \delta \left(\frac{k_{L,1}^2}{E_1}\right) \right],
\end{equation}
and working to first order in small quantities 
\begin{equation}
\label{eq:delc}
\delta C =\frac{3}{Q}  \left [E_2- \frac{k_{L,2}^2}{E_2} +\delta E_1 +\frac{k_{L,1}^2}{E_1} \left(\frac{\delta E_1}{E_1}-2\frac{\delta k_{L,1}}{k_{L,1}} \right)\right],
\end{equation}
which implies that we simply need to evaluate the change in the longitudinal momentum and energy of emission $k_1$ due to the emission of $k_2$. We now evaluate this for various situations i.e. for the case of various emitting dipoles, and for each dipole the cases where one or the other leg is designated as emitter or spectator.
Similar to the thrust case we therefore write:
\begin{equation}
\label{eq:pqc1}
\delta C^{p_i,p_j} = \delta C^{p_i p_j,\,p_i \,\, \text{emitter}} \theta \left( \theta_{2i}< \theta_{2j} \right)+ \delta C^{p_i p_j,\,p_j \,\, \text{emitter}} \theta \left( \theta_{2i}> \theta_{2j} \right).
\end{equation}
We evaluate the various dipole contributions below.
\begin{itemize}
\item $p_q k_1$ dipole:\\
The overall result for this dipole may be expressed in terms of the cases for when $k_1$ is the emitter or the spectator. When $k_1$ is the spectator it only takes longitudinal recoil. Thus after emission of $k_2$ we have $k_1= y\tilde{k}_1 $ and $\delta E_1/E_1 =\delta k_{L,1}/k_{L,1}=-(1-y)$. From Eq.~\eqref{eq:delc} we can write 
\begin{equation}
\label{eq:delc1}
\delta C^{p_q k_1, \, p_q \,\, \text{emitter}} = \frac{3}{Q} \left(E_2- \frac{k_{L,2}^2}{E_2} -(1-y) \left(E_1- \frac{k_{L,1}^2}{E_1} \right) \right).
\end{equation}

The quantity $y$ is given in our dipole local recoil scheme by 
\begin{equation}
y = \beta_{k1} \approx 1- \frac{p_q \cdot k_2}{p_q \cdot k_1},
\end{equation}
where we worked to first order in the softest emission $k_2$ which allows us to replace $\tilde{p}_q, \tilde{k}_1$ by $p_q$ and $k_1$ respectively in obtaining $y$. In terms of our usual kinematic variables we obtain:
\begin{equation}
\label{eq:delc2}
\delta C^{p_q k_1, \, p_q \,\, \text{emitter}} = \frac{3}{2} x_2 \,  (1-\cos \theta_{2q})(\cos\theta_{2q}-\cos\theta_{1q}).
\end{equation}

Now we handle the case where $k_1$ is the emitter and $p_q$ the spectator. This follows similarly except that now $p_q$ takes only longitudinal recoil i.e. $p_q \to y \tilde{p}_q$, with $y =\beta_q \approx 1-\frac{k_1 \cdot k_2}{p_q \cdot k_1}$.
The above observation alongside straightforward energy-momentum conservation allows us to work out all the quantities involved in eq.~\eqref{eq:delc}. We get:
\begin{equation}
\label{eq:delc3}
\delta C^{p_q k_1, \, k_1 \,\, \text{emitter}} = \frac{3}{2} x_2 \left[ (1-\cos\theta_{12})(1-\cos \theta_{1q}) -(\cos\theta_{2q} -\cos \theta_{1q})^2 \right].
\end{equation}

\item $\pqbar k_1$ dipole: \\
Following the same considerations as for the $p_q k_1$ dipole it is straightforward to derive the results:
\begin{align}
\label{eq:delc4}
\delta C^{\pqbar k_1, \, \pqbar \,\, \text{emitter}} &= \frac{3}{2} x_2 \,  (1+\cos \theta_{2q})(\cos\theta_{1q}-\cos\theta_{2q}) ,\\ 
\delta C^{\pqbar k_1, \, k_1 \,\, \text{emitter}} &= \frac{3}{2} x_2 \left[ (1-\cos\theta_{12})(1+\cos \theta_{1q}) -(\cos\theta_{2q} -\cos \theta_{1q})^2 \right].
\end{align}
\item $p_q \pqbar$ dipole: \\
The results for this dipole are simple to obtain since $k_1$ does not take recoil. So the situation is the same as the independent emission from a $q\bar{q}$ dipole case and the result obtained, to first order in the softness of $k_2$,  is simply the contribution of emission $k_2$ to the $C$-parameter given by:
\begin{equation}
\label{eq:delc5}
\delta C^{p_q \pqbar} = \frac{3}{2} x_2 \sin^2 \theta_{2q}. 
\end{equation}
We note that here there is no need to explicitly separate the emitter and spectator cases, as a unique answer is obtained.
\end{itemize}
\subsection{$C$-parameter integrals and result}
The evaluation of the result for the $C$-parameter now proceeds identically to that for the thrust and we detail the different numerical contributions below. 
\subsubsection{Emission from the $p_q k_1$ dipole}
For the $p_q k_1$ dipole for instance we obtain the result analogous to eq.~\eqref{eq:p1k1int} for the thrust by evaluating a similar integral differing only in the functional form of the observable:
\begin{equation}
\label{eq:p1k1intc}
 \langle  \delta C \rangle^{\text{NP},1,p_q k_1} =  8 C_F C_A \frac{\alpha_s}{2\pi} \int_0^{\mu_I} d\kappa_T \frac{\alpha_s(\kappa_T)}{2\pi} \ln \frac{Q}{\kappa_T} \times I_C^{p_q k_1},
\end{equation}

where
\begin{equation}
I_C^{p_q k_1} =  \int_{0}^{1} d\cos{\theta_{1q}} \int_{-1}^1 d\cos{\theta_{2q}}\int_0^{2\pi} \frac{d\phi}{2\pi}  \frac{\sqrt{2} \sqrt{1-\cos\theta_{1q}}}{\left(1-\cos\theta_{12} \right)^{3/2} (1+\cos\theta_{1q})(1-\cos\theta_{2q})^{3/2}} \times \delta C^{p_q k_1},
\end{equation}
with
\begin{equation}
\delta C^{p_q k_1} = \delta C^{p_q k_1, \, \pq \,\, \text{emitter}}\theta \left( \theta_{2q}< \theta_{12} \right)+ \delta C^{p_q k_1, \, k_1 \,\, \text{emitter}}\theta \left( \theta_{2q}> \theta_{12} \right).
\end{equation}
Using eqs.~\eqref{eq:delc2} and \eqref{eq:delc3} and evaluating the integral we get \footnote{The numerical accuracy we are able to obtain for the $C$-parameter, while at a per-mille error level, is still considerably less than the corresponding results for the case of the thrust.}

\begin{equation}
I_C^{p_q k_1} \approx 0.830 \pm 0.002.
\end{equation}

\subsubsection{Emission from the $p_{\bar{q} }k_1$ and $p_q \pqbar$ dipoles}
For the $\pqbar k_1$ dipole, to handle efficiently the cancellation of collinear divergences, we proceed as for the thrust case. To be precise we consider a simplified version of the calculation where if the emission $k_2$ is in  $\mathcal{H}_L$ we take $\pqbar$ as emitter, otherwise $k_1$ is designated the emitter. As before we shall combine the $k_2 \in \mathcal{H}_L$ and  $k_2 \in \mathcal{H}_R$ cases with the corresponding regions for the $p_q \pqbar$ dipole term and compute separate correction terms for each contribution that account for the difference between the true and simplified recoil assignments. Doing so gives rise to the following two extra contributions. Firstly for the case where $k_2 \in \mathcal{H_R}$ we get \footnote{We remind the reader that we place $k_1$ in $\mathcal{H}_R$ and account for the other hemisphere via a factor of two.}

\begin{multline}
\label{eq:csimp1}
I_C^{\pqbar k_1, \text{simp.},k_2 \in \mathcal{H}_R}+I_C^{p_q \pqbar, k_2 \in \mathcal{H}_R}  = \int_{0}^{1} d\cos{\theta_{1q}} \int_{0}^1 d\cos{\theta_{2q}}\int_0^{2\pi} \frac{d\phi}{2\pi} \times \\ \left [ \frac{\sqrt{2}\sqrt{1+\cos\theta_{1q}}}{(1+\cos\theta_{2q})^{3/2} (1-\cos\theta_{1q}) \left(1-\cos\theta_{12} \right)^{3/2}} \delta C^{\pqbar k_1, \, k_1 \,\, \text{emitter}} \right .\\
\left .-\frac{4}{(1-\cos^2\theta_{1q}) (1-\cos^2\theta_{2q})^{3/2}} \delta C^{p_q \pqbar} \right],
\end{multline}
and on performing the integrals numerically we find:
\begin{equation}
I_C^{\pqbar k_1, \text{simp.},k_2 \in \mathcal{H}_R}+I_C^{p_q \pqbar, k_2 \in \mathcal{H}_R}  =  -7.015 \pm 0.007.
\end{equation}

The case of $k_2 \in \mathcal{H}_L$ is handled similarly with the only changes needed to eq.~\eqref{eq:csimp1} being the use of $\delta C^{\pqbar k_1, \, \pqbar \,\, \text{emitter}}$ instead of $ \delta C^{\pqbar k_1, \, k_1 \,\, \text{emitter}}$ and the fact that the $\cos \theta_{2q}$ integration now extends from $-1$ to $0$. On numerical evaluation this gives the result

\begin{equation}
I_C^{\pqbar k_1, \text{simp.},k_2 \in \mathcal{H}_L}+I_C^{p_q \pqbar, k_2 \in \mathcal{H}_L}  =  -1.468 \pm 0.003.
\end{equation}
Finally we account for the correction terms that are needed to correct our simplified treatment of the $\pqbar k_1$ dipole over the region where there is a difference between the true and simplified situations. For the case when $k_2 \in \mathcal{H}_R$, we have to evaluate 

\begin{multline}
\label{eq:ccorr1}
 I_C^{\pqbar k_1, \text{corr.},k_2 \in \mathcal{H}_R} =   \int_{0}^{1} d\cos{\theta_{1q}} \int_{0}^{1} d\cos{\theta_{2q}}\int_0^{2\pi} \frac{d\phi}{2\pi}  \frac{\sqrt{2}\sqrt{1+\cos\theta_{1q}}}{(1+\cos\theta_{2q})^{3/2} (1-\cos\theta_{1q}) \left(1-\cos\theta_{12} \right)^{3/2}} \times \\ \times \left[ \delta C^{\pqbar k_1, \, \pqbar \,\, \text{emitter}}-\delta C^{\pqbar k_1, \, k_1 \,\, \text{emitter}} \right]  \theta(\theta_{2\bar{q}}<\theta_{12}),
\end{multline}
which we evaluate numerically to obtain:
\begin{equation}
I_C^{\pqbar k_1, \text{corr.},k_2 \in \mathcal{H}_R} = -0.695 \pm 0.0007.
\end{equation}

Finally we have to evaluate a similar correction for the case when $k_2 \in \mathcal{H}_L$. Here in the simplified treatment the emitter was $\pqbar$ while over a sub-region when $k_2$ is closer to $k_1$ the emitter should be $k_1$.
The integral to evaluate is the same as in \eqref{eq:ccorr1} except that in the second line we use the factor $\left[ \delta C^{\pqbar k_1, \, k_1\,\, \text{emitter}}-\delta C^{\pqbar k_1, \, \pqbar \,\, \text{emitter}} \right]  \theta(\theta_{2\bar{q}} > \theta_{12})$ and the limits on the $\cos \theta_{2q}$ extend from -1 to 0. The result is
\begin{equation}
I_C^{\pqbar k_1, \text{corr.},k_2 \in \mathcal{H}_R} = -3.222 \pm 0.003.
\end{equation}

Combining all terms from our numerical evaluations and multiplying by the relevant prefactor, we obtain the result analogous to that for the thrust reported in eq.~\eqref{eq:thrustres}:
\begin{equation}
\label{eq:cres}
 \langle  \delta C \rangle^{\text{NP},1} =   -C_F C_A  \frac{\alpha_s}{2\pi}  \frac{1}{Q} \int_0^{\mu_I} d\kappa_T \frac{\alpha_s(\kappa_T)}{2\pi} \ln \frac{Q}{\kappa_T} \times \left(92.56 \pm 0.126 \right).
\end{equation}

\subsection{Comparison to thrust, universality of power corrections}
A well-known feature of $1/Q$ power corrections to linear event shape variables is that of {\it{universality}}. Specifically this means that the relative values of power corrections to the event shape variables that share this property are governed by a simple one-gluon calculation. Event shapes of this type receive a universal two-loop enhancement factor (Milan factor) and the non-perturbative physics is contained within the same $\alpha_0$ parameter for all event shapes. Thus ratios of power corrections to event shape mean values, for example, come out simply as the ratio of the coefficients computed in the standard DW model. For $\langle \tau \rangle$ and $\langle C \rangle$ considered here, the ratio of the standard DW results is given by 
$\frac{3\pi}{2}$, see Eq.~\eqref{eq:loc}. This ratio is maintained even after computing the impact of gluon decay via the Milan factor, owing to its universality. 

It is therefore interesting to see if the universality picture holds for the $Q$ evolution i.e. also for the anomalous dimension. We thus consider the ratio of the $C$-parameter result to that we obtained for the thrust, for the first-order in $\alpha_s$ contribution we have computed here. We get:
\begin{equation}
\frac{ \langle  \delta C \rangle^{\text{NP},1} }{ \langle  \delta \tau \rangle^{\text{NP},1} } =4.712 \pm 0.006 \approx \frac{3\pi}{2},
\end{equation}
where we note that $3\pi/2 =4.712713\cdots$. Hence to our numerical accuracy we obtain a result where the ratio of the $C$-parameter to the thrust results is the same as the ratio of the results obtained within the single-gluon approximation using the standard Dokshitzer-Webber model. The above result indicates that the anomalous dimension terms preserve the key feature of universality of power corrections for event shape variables. This finding could perhaps have been anticipated. The universality of the Milan factor comes out owing to the boost invariance along the $q\bar{q}$ axis enjoyed by the double-soft matrix-element and the linearity of the event shape in terms of the final state momenta. Both these properties are also relevant here. Nevertheless we performed the calculation for the thrust and C-parameter independently and in full, which also acts as a check on our calculations.

We also note that while the current article was being prepared, we were made aware of another ongoing study investigating the question of anomalous dimensions in $1/Q$ corrections to event shapes \cite{gavin}. Results emerging from that study coincide with ours in the context of the leading-order calculations performed here. \footnote{We thank the authors of Ref.~\cite{gavin} for informing us about this finding.}

\section{Discussion and conclusions}
\label{sec:disc}
The work we have presented here highlights a somewhat neglected aspect of non-perturbative $1/Q$ corrections, which we believe is already of phenomenological relevance. We have demonstrated how to compute, via a simple extension of the standard Dokshitzer-Webber model, a perturbative logarithmic enhancement of the leading $1/Q$ power correction for event shapes. Our results show that the usual $1/Q$ correction of the DW model, as first computed in Ref.~\cite{Dokshitzer_1995}, receives logarithmically enhanced higher order corrections such that for a two-jet event shape variable $v$  in $e^{+}e^{-}$ annihilation one has a result of the form:
\begin{equation}
\label{eq:finalad}
\langle \delta v \rangle^{\text{NP}}  = C_{v} \frac{2 C_F }{\pi Q} \int_0^{\mu_I} \alpha_s(k_\perp) dk_\perp \left( 1-\mathcal{S}_1 C_A \frac{\alpha_s}{2\pi}  \ln\frac{Q}{k_\perp} +\cdots\right),
\end{equation}
where $C_v=2$ for thrust and $C_v=3 \pi$ for the $C$-parameter. Using our results from this article, we find $\mathcal{S}_1 \approx 2.455$. The ellipses denote as yet undetermined higher order terms of the form $\alpha_s^n L^n$ with $L = \ln Q/k_\perp$ and $n \geq 2$, and the series of terms in parenthesis would need to be resummed, prior to any application.

To bring Eq.~\eqref{eq:finalad} into the standard form used for event shape studies one needs to multiply by a factor $\frac{2}{\pi} \mathcal{M}$ where $\mathcal{M}=1.49 \, (n_f=3)$ is the universal Milan factor.\footnote{Although we have not computed it explicitly in the present context, the Milan factor arises from a careful consideration of gluer decay and should also apply to the full anomalous dimension term.} 
The subtraction of the perturbative contribution to the $\alpha_s$ integral, up to order $\alpha_s^2$ accuracy may be applied as usual. We refer the reader to Ref.~\cite{Dasgupta:2003iq} for a detailed discussion of these points, as applied to the original form of $1/Q$ power corrections. Here we continue to focus on the logarithmic $Q$ dependence we have computed, where a number of further developments can be envisaged. 

Firstly the resummation of the perturbative series in parenthesis should be carried out to determine the full functional form at leading-logarithmic accuracy and the true size of the effect, since here we have limited ourselves to just computing the leading $\mathcal{O}(\alpha_s)$ term. Such a resummation would involve addressing gluer emission from an ensemble of soft commensurate-angle 
perturbative emissions ordered in energy or equivalently $k_\perp$, and hence resembles closely the resummation of non-global logarithms. While we have carried out our fixed-order calculation in this paper with full dependence on colour retained, the non-global resummation required is most conveniently carried out in a leading $N_c$ approximation, though full $N_c$ non-global approaches also exist \cite{Hatta:2013iba}. 

With the resummation in hand it would be important to understand the phenomenological relevance of the effect discussed here. For event shape mean values the $Q$ dependent coefficient 
of the power corrections would be a factor in fits of power corrections, leading to a modification of the standard $\alpha_0$ parameter extractions.\footnote{As a rather rough estimate one can replace the argument of the logarithm by $Q/\mu_I$ in Eq.~\eqref{eq:finalad} and assume a simple exponentiation of our result. Doing so and using $\alpha_s=0.118$, $Q=90$ GeV, and $\mu_I = 2$ GeV, suggests that a value of $\alpha_0 \approx 0.4$ should, once the anomalous dimension is included, be consistent with a revised value $\alpha_0 \approx 0.68$.}
The study of the impact on differential distributions for event shapes in $e^{+}e^{-}$ annihilation would also be important to explore. Another area of applicability is to power corrections for small $R$ jets at the LHC \cite{Dasgupta:2007wa} and related calculations for jet substructure observables such as groomed jet masses \cite{Dasgupta:2013ihk,Hoang:2019ceu}.  The power correction to the average shift in a small $R$ jet's transverse momentum was found in Ref.~\cite{Dasgupta:2007wa} to take the form of a $1/R$ power correction, with a coefficient which is one-half of that for the $e^{+}e^{-}$ thrust. The calculations of Refs.~\cite{Dasgupta:2007wa} were carried out in a single gluon DW framework. Beyond the single-gluon level however, jet clustering effects kick in and spoil the simple picture of universality, in that one needs to compute a Milan factor that depends on the jet algorithm \cite{Dasgupta:2009tm,Hounat:2023nvc}. Such a non-universality might also apply for the anomalous dimension, which might need recomputation for different algorithms.\footnote{A possible exception might be the anti-$k_t$ case which is more resilient to soft gluon clustering effects \cite{Cacciari:2008gp}.} Nevertheless the anomalous dimension should give a logarithmic $p_t$ dependence to the $1/R$ correction. A dependence of the hadronization correction on $p_t$, in the case of the inclusive jet spectrum was in fact noted in the Monte Carlo studies of hadronization carried out in Ref.~\cite{Dasgupta:2016bnd}. However one would need to include the perturbative evolution effect of this paper alongside other possible sources of such logarithmic $p_t$ dependence, notably due to hadron mass effects \cite{Salam:2001bd} prior to making phenomenological statements. Finally the additional scale dependence of the $1/Q$ power corrections, identified here, should also be kept in mind when carrying out studies of power corrections at possible future colliders and comparing the phenomenology to that for event shapes at LEP. 

Finally, in terms of comparison to alternative approaches to non-perturbative power corrections, it would be useful to study the $Q$ dependence of hadronization corrections in the various hadronization models of Monte Carlo event generators, and compare the findings with the anomalous dimension emerging from a full resummed extension of our initial study.\footnote{Studies along the lines of those carried out in Ref.~\cite{Hoang:2024zwl} might also be relevant to such investigations.} On a more formal note it would also be desirable to establish a deeper connection with the renormalon picture that inspired many of the developments regarding analytical hadronization models \cite{Beneke:1998ui}. Such a connection is however non-trivial, since the effect we have computed in the present article would not appear in a large $n_f$ abelian limit, which is the strict formal domain of validity of the renormalon calculus. It might also be natural to think of the anomalous dimension as being associated to perturbative evolution in an operator language for $1/Q$ corrections (see e.g. Ref.~\cite{Korchemsky:1997sy}). We shall leave an exploration of some of these topics to future work.

\section*{Acknowledgments}
One of us (MD) is grateful to Gavin Salam for numerous useful discussions and sharing insight related to the topic of this paper. We are also grateful to the authors of Ref.~\cite{gavin} (Fabrizio Caola, Casey Farren-Colloty, Jack Helliwell, Rtvik Patel, Gavin P. Salam, Silvia Zanoli) for informing us of their related ongoing work, and for sharing results showing that their study gives results consistent with those presented here.
We also thank Pier Monni for interesting discussions on higher-order power corrections and non-global resummation and are grateful to the CERN theory department for hospitality while this work was in progress. We are grateful to the U.K.'s STFC for financial support via grants ST/T001038/1, ST/00077X/1 (MD) and a PhD studentship award (FH).

\bibliographystyle{JHEP}
\bibliography{powercorr}

\providecommand{\href}[2]{#2}\begingroup\raggedright\begin{thebibliography}{100}

\bibitem{ref}
{\scshape FCC} collaboration, A.~Abada et~al., \emph{{FCC-ee: The Lepton
  Collider}: {Future Circular Collider Conceptual Design Report Volume 2}},
  \href{https://doi.org/10.1140/epjst/e2019-900045-4}{\emph{Eur. Phys. J. ST}
  {\bfseries 228} (2019) 261}.

\bibitem{Gehrmann-DeRidder:2007nzq}
A.~Gehrmann-De~Ridder, T.~Gehrmann, E.~W.~N. Glover and G.~Heinrich,
  \emph{{Second-order QCD corrections to the thrust distribution}},
  \href{https://doi.org/10.1103/PhysRevLett.99.132002}{\emph{Phys. Rev. Lett.}
  {\bfseries 99} (2007) 132002}
  [\href{https://arxiv.org/abs/0707.1285}{{\ttfamily 0707.1285}}].

\bibitem{Dissertori:2008cn}
G.~Dissertori, A.~Gehrmann-De~Ridder, T.~Gehrmann, E.~W.~N. Glover, G.~Heinrich
  and H.~Stenzel, \emph{{$e^+e^-$ to 3 jets and event shapes at NNLO}},
  \href{https://doi.org/10.1016/j.nuclphysbps.2008.09.072}{\emph{Nucl. Phys. B
  Proc. Suppl.} {\bfseries 183} (2008) 2}
  [\href{https://arxiv.org/abs/0806.4601}{{\ttfamily 0806.4601}}].

\bibitem{Gehrmann-DeRidder:2009fgd}
A.~Gehrmann-De~Ridder, T.~Gehrmann, E.~W.~N. Glover and G.~Heinrich,
  \emph{{NNLO moments of event shapes in $e^+e^-$ annihilation}},
  \href{https://doi.org/10.1088/1126-6708/2009/05/106}{\emph{JHEP} {\bfseries
  05} (2009) 106} [\href{https://arxiv.org/abs/0903.4658}{{\ttfamily
  0903.4658}}].

\bibitem{Weinzierl:2008iv}
S.~Weinzierl, \emph{{NNLO corrections to 3-jet observables in electron-positron
  annihilation}},
  \href{https://doi.org/10.1103/PhysRevLett.101.162001}{\emph{Phys. Rev. Lett.}
  {\bfseries 101} (2008) 162001}
  [\href{https://arxiv.org/abs/0807.3241}{{\ttfamily 0807.3241}}].

\bibitem{Weinzierl_2009yz}
S.~Weinzierl, \emph{Event shapes and jet rates in electron-positron
  annihilation at {NNLO}},
  \href{https://doi.org/10.1088/1126-6708/2009/06/041}{\emph{Journal of High
  Energy Physics} {\bfseries 2009} (2009) 041}.

\bibitem{Gehrmann:2019hwf}
T.~Gehrmann, A.~Huss, J.~Mo and J.~Niehues, \emph{{Second-order QCD corrections
  to event shape distributions in deep inelastic scattering}},
  \href{https://doi.org/10.1140/epjc/s10052-019-7528-3}{\emph{Eur. Phys. J. C}
  {\bfseries 79} (2019) 1022}
  [\href{https://arxiv.org/abs/1909.02760}{{\ttfamily 1909.02760}}].

\bibitem{Alvarez:2023fhi}
M.~Alvarez, J.~Cantero, M.~Czakon, J.~Llorente, A.~Mitov and R.~Poncelet,
  \emph{{NNLO QCD corrections to event shapes at the LHC}},
  \href{https://doi.org/10.1007/JHEP03(2023)129}{\emph{JHEP} {\bfseries 03}
  (2023) 129} [\href{https://arxiv.org/abs/2301.01086}{{\ttfamily
  2301.01086}}].

\bibitem{Ebert:2020sfi}
M.~A. Ebert, B.~Mistlberger and G.~Vita, \emph{{The Energy-Energy Correlation
  in the back-to-back limit at N$^{3}$LO and N$^{3}$LL'}},
  \href{https://doi.org/10.1007/JHEP08(2021)022}{\emph{JHEP} {\bfseries 08}
  (2021) 022} [\href{https://arxiv.org/abs/2012.07859}{{\ttfamily
  2012.07859}}].

\bibitem{Bozzi:2005wk}
G.~Bozzi, S.~Catani, D.~de~Florian and M.~Grazzini, \emph{{Transverse-momentum
  resummation and the spectrum of the Higgs boson at the LHC}},
  \href{https://doi.org/10.1016/j.nuclphysb.2005.12.022}{\emph{Nucl. Phys. B}
  {\bfseries 737} (2006) 73}
  [\href{https://arxiv.org/abs/hep-ph/0508068}{{\ttfamily hep-ph/0508068}}].

\bibitem{Bozzi:2010xn}
G.~Bozzi, S.~Catani, G.~Ferrera, D.~de~Florian and M.~Grazzini,
  \emph{{Production of Drell-Yan lepton pairs in hadron collisions:
  Transverse-momentum resummation at next-to-next-to-leading logarithmic
  accuracy}}, \href{https://doi.org/10.1016/j.physletb.2010.12.024}{\emph{Phys.
  Lett. B} {\bfseries 696} (2011) 207}
  [\href{https://arxiv.org/abs/1007.2351}{{\ttfamily 1007.2351}}].

\bibitem{Stewart_2011}
I.~W. Stewart, F.~J. Tackmann and W.~J. Waalewijn, \emph{Beam thrust cross
  section for drell-yan production at next-to-next-to-leading-logarithmic
  order}, \href{https://doi.org/10.1103/physrevlett.106.032001}{\emph{Physical
  Review Letters} {\bfseries 106} (2011) }.

\bibitem{Banfi:2011dm}
A.~Banfi, M.~Dasgupta, S.~Marzani and L.~Tomlinson, \emph{{Probing the low
  transverse momentum domain of Z production with novel variables}},
  \href{https://doi.org/10.1007/JHEP01(2012)044}{\emph{JHEP} {\bfseries 01}
  (2012) 044} [\href{https://arxiv.org/abs/1110.4009}{{\ttfamily 1110.4009}}].

\bibitem{Banfi:2012jm}
A.~Banfi, P.~F. Monni, G.~P. Salam and G.~Zanderighi, \emph{{Higgs and Z-boson
  production with a jet veto}},
  \href{https://doi.org/10.1103/PhysRevLett.109.202001}{\emph{Phys. Rev. Lett.}
  {\bfseries 109} (2012) 202001}
  [\href{https://arxiv.org/abs/1206.4998}{{\ttfamily 1206.4998}}].

\bibitem{Banfi:2015pju}
A.~Banfi, F.~Caola, F.~A. Dreyer, P.~F. Monni, G.~P. Salam, G.~Zanderighi
  et~al., \emph{{Jet-vetoed Higgs cross section in gluon fusion at
  N$^{3}$LO+NNLL with small-$R$ resummation}},
  \href{https://doi.org/10.1007/JHEP04(2016)049}{\emph{JHEP} {\bfseries 04}
  (2016) 049} [\href{https://arxiv.org/abs/1511.02886}{{\ttfamily
  1511.02886}}].

\bibitem{Jouttenus_2013}
T.~T. Jouttenus, I.~W. Stewart, F.~J. Tackmann and W.~J. Waalewijn, \emph{Jet
  mass spectra in higgs boson plus one jet at next-to-next-to-leading
  logarithmic order},
  \href{https://doi.org/10.1103/physrevd.88.054031}{\emph{Physical Review D}
  {\bfseries 88} (2013) }.

\bibitem{Kang:2013nha}
D.~Kang, C.~Lee and I.~W. Stewart, \emph{{Using 1-Jettiness to Measure 2 Jets
  in DIS 3 Ways}},
  \href{https://doi.org/10.1103/PhysRevD.88.054004}{\emph{Phys. Rev. D}
  {\bfseries 88} (2013) 054004}
  [\href{https://arxiv.org/abs/1303.6952}{{\ttfamily 1303.6952}}].

\bibitem{Banfi:2014sua}
A.~Banfi, H.~McAslan, P.~F. Monni and G.~Zanderighi, \emph{{A general method
  for the resummation of event-shape distributions in $e^{+} e^{-}$
  annihilation}}, \href{https://doi.org/10.1007/JHEP05(2015)102}{\emph{JHEP}
  {\bfseries 05} (2015) 102} [\href{https://arxiv.org/abs/1412.2126}{{\ttfamily
  1412.2126}}].

\bibitem{Becher:2015gsa}
T.~Becher and X.~Garcia~i Tormo, \emph{{Factorization and resummation for
  transverse thrust}},
  \href{https://doi.org/10.1007/JHEP06(2015)071}{\emph{JHEP} {\bfseries 06}
  (2015) 071} [\href{https://arxiv.org/abs/1502.04136}{{\ttfamily
  1502.04136}}].

\bibitem{Banfi:2016zlc}
A.~Banfi, H.~McAslan, P.~F. Monni and G.~Zanderighi, \emph{{The two-jet rate in
  $e^+e^-$ at next-to-next-to-leading-logarithmic order}},
  \href{https://doi.org/10.1103/PhysRevLett.117.172001}{\emph{Phys. Rev. Lett.}
  {\bfseries 117} (2016) 172001}
  [\href{https://arxiv.org/abs/1607.03111}{{\ttfamily 1607.03111}}].

\bibitem{Becher_2016}
T.~Becher, X.~G. i~Tormo and J.~Piclum, \emph{Next-to-next-to-leading
  logarithmic resummation for transverse thrust},
  \href{https://doi.org/10.1103/physrevd.93.054038}{\emph{Physical Review D}
  {\bfseries 93} (2016) }.

\bibitem{Tulipant:2017ybb}
Z.~Tulip\'ant, A.~Kardos and G.~Somogyi, \emph{{Energy\textendash{}energy
  correlation in electron\textendash{}positron annihilation at NNLL + NNLO
  accuracy}}, \href{https://doi.org/10.1140/epjc/s10052-017-5320-9}{\emph{Eur.
  Phys. J. C} {\bfseries 77} (2017) 749}
  [\href{https://arxiv.org/abs/1708.04093}{{\ttfamily 1708.04093}}].

\bibitem{Banfi:2018mcq}
A.~Banfi, B.~K. El-Menoufi and P.~F. Monni, \emph{{The Sudakov radiator for jet
  observables and the soft physical coupling}},
  \href{https://doi.org/10.1007/JHEP01(2019)083}{\emph{JHEP} {\bfseries 01}
  (2019) 083} [\href{https://arxiv.org/abs/1807.11487}{{\ttfamily
  1807.11487}}].

\bibitem{Gao:2019ojf}
A.~Gao, H.~T. Li, I.~Moult and H.~X. Zhu, \emph{{Precision QCD Event Shapes at
  Hadron Colliders: The Transverse Energy-Energy Correlator in the Back-to-Back
  Limit}}, \href{https://doi.org/10.1103/PhysRevLett.123.062001}{\emph{Phys.
  Rev. Lett.} {\bfseries 123} (2019) 062001}
  [\href{https://arxiv.org/abs/1901.04497}{{\ttfamily 1901.04497}}].

\bibitem{Kardos:2020gty}
A.~Kardos, A.~J. Larkoski and Z.~Tr\'ocs\'anyi, \emph{{Groomed jet mass at high
  precision}},
  \href{https://doi.org/10.1016/j.physletb.2020.135704}{\emph{Phys. Lett. B}
  {\bfseries 809} (2020) 135704}
  [\href{https://arxiv.org/abs/2002.00942}{{\ttfamily 2002.00942}}].

\bibitem{Dasgupta:2022fim}
M.~Dasgupta, B.~K. El-Menoufi and J.~Helliwell, \emph{{QCD resummation for
  groomed jet observables at NNLL+NLO}},
  \href{https://doi.org/10.1007/JHEP01(2023)045}{\emph{JHEP} {\bfseries 01}
  (2023) 045} [\href{https://arxiv.org/abs/2211.03820}{{\ttfamily
  2211.03820}}].

\bibitem{vanBeekveld:2023lsa}
M.~van Beekveld, M.~Dasgupta, B.~K. El-Menoufi, J.~Helliwell and P.~F. Monni,
  \emph{{Collinear fragmentation at NNLL: generating functionals, groomed
  correlators and angularities}},
  \href{https://arxiv.org/abs/2307.15734}{{\ttfamily 2307.15734}}.

\bibitem{Chen:2023zlx}
W.~Chen, J.~Gao, Y.~Li, Z.~Xu, X.~Zhang and H.~X. Zhu, \emph{{NNLL Resummation
  for Projected Three-Point Energy Correlator}},
  \href{https://arxiv.org/abs/2307.07510}{{\ttfamily 2307.07510}}.

\bibitem{Bhattacharya:2023qet}
A.~Bhattacharya, J.~K.~L. Michel, M.~D. Schwartz, I.~W. Stewart and X.~Zhang,
  \emph{{NNLL Resummation of Sudakov Shoulder Logarithms in the Heavy Jet Mass
  Distribution}},  \href{https://arxiv.org/abs/2306.08033}{{\ttfamily
  2306.08033}}.

\bibitem{Becher:2008cf}
T.~Becher and M.~D. Schwartz, \emph{{A precise determination of $\alpha_s$ from
  LEP thrust data using effective field theory}},
  \href{https://doi.org/10.1088/1126-6708/2008/07/034}{\emph{JHEP} {\bfseries
  07} (2008) 034} [\href{https://arxiv.org/abs/0803.0342}{{\ttfamily
  0803.0342}}].

\bibitem{Chien:2010kc}
Y.-T. Chien and M.~D. Schwartz, \emph{{Resummation of heavy jet mass and
  comparison to LEP data}},
  \href{https://doi.org/10.1007/JHEP08(2010)058}{\emph{JHEP} {\bfseries 08}
  (2010) 058} [\href{https://arxiv.org/abs/1005.1644}{{\ttfamily 1005.1644}}].

\bibitem{Abbate:2010xh}
R.~Abbate, M.~Fickinger, A.~H. Hoang, V.~Mateu and I.~W. Stewart, \emph{{Thrust
  at $N^{3}LL$ with Power Corrections and a Precision Global Fit for
  $\alpha_{s}(mZ)$}},
  \href{https://doi.org/10.1103/PhysRevD.83.074021}{\emph{Phys. Rev. D}
  {\bfseries 83} (2011) 074021}
  [\href{https://arxiv.org/abs/1006.3080}{{\ttfamily 1006.3080}}].

\bibitem{Hoang:2014wka}
A.~H. Hoang, D.~W. Kolodrubetz, V.~Mateu and I.~W. Stewart,
  \emph{{$C$-parameter distribution at N$^3$LL' including power corrections}},
  \href{https://doi.org/10.1103/PhysRevD.91.094017}{\emph{Phys. Rev. D}
  {\bfseries 91} (2015) 094017}
  [\href{https://arxiv.org/abs/1411.6633}{{\ttfamily 1411.6633}}].

\bibitem{Forshaw:2019ver}
J.~R. Forshaw, J.~Holguin and S.~Pl\"atzer, \emph{{Parton branching at
  amplitude level}}, \href{https://doi.org/10.1007/JHEP08(2019)145}{\emph{JHEP}
  {\bfseries 08} (2019) 145}
  [\href{https://arxiv.org/abs/1905.08686}{{\ttfamily 1905.08686}}].

\bibitem{Forshaw:2020wrq}
J.~R. Forshaw, J.~Holguin and S.~Pl\"atzer, \emph{{Building a consistent parton
  shower}}, \href{https://doi.org/10.1007/JHEP09(2020)014}{\emph{JHEP}
  {\bfseries 09} (2020) 014}
  [\href{https://arxiv.org/abs/2003.06400}{{\ttfamily 2003.06400}}].

\bibitem{Dasgupta:2020fwr}
M.~Dasgupta, F.~A. Dreyer, K.~Hamilton, P.~F. Monni, G.~P. Salam and G.~Soyez,
  \emph{{Parton showers beyond leading logarithmic accuracy}},
  \href{https://doi.org/10.1103/PhysRevLett.125.052002}{\emph{Phys. Rev. Lett.}
  {\bfseries 125} (2020) 052002}
  [\href{https://arxiv.org/abs/2002.11114}{{\ttfamily 2002.11114}}].

\bibitem{Hamilton:2020rcu}
K.~Hamilton, R.~Medves, G.~P. Salam, L.~Scyboz and G.~Soyez, \emph{{Colour and
  logarithmic accuracy in final-state parton showers}},
  \href{https://doi.org/10.1007/JHEP03(2021)041}{\emph{JHEP} {\bfseries 03}
  (2021) 041} [\href{https://arxiv.org/abs/2011.10054}{{\ttfamily
  2011.10054}}].

\bibitem{Nagy:2020dvz}
Z.~Nagy and D.~E. Soper, \emph{{Summations by parton showers of large
  logarithms in electron-positron annihilation}},
  \href{https://arxiv.org/abs/2011.04777}{{\ttfamily 2011.04777}}.

\bibitem{Nagy:2020rmk}
Z.~Nagy and D.~E. Soper, \emph{{Summations of large logarithms by parton
  showers}}, \href{https://doi.org/10.1103/PhysRevD.104.054049}{\emph{Phys.
  Rev. D} {\bfseries 104} (2021) 054049}
  [\href{https://arxiv.org/abs/2011.04773}{{\ttfamily 2011.04773}}].

\bibitem{Karlberg:2021kwr}
A.~Karlberg, G.~P. Salam, L.~Scyboz and R.~Verheyen, \emph{{Spin correlations
  in final-state parton showers and jet observables}},
  \href{https://doi.org/10.1140/epjc/s10052-021-09378-0}{\emph{Eur. Phys. J. C}
  {\bfseries 81} (2021) 681}
  [\href{https://arxiv.org/abs/2103.16526}{{\ttfamily 2103.16526}}].

\bibitem{Hamilton:2021dyz}
K.~Hamilton, A.~Karlberg, G.~P. Salam, L.~Scyboz and R.~Verheyen, \emph{{Soft
  spin correlations in final-state parton showers}},
  \href{https://doi.org/10.1007/JHEP03(2022)193}{\emph{JHEP} {\bfseries 03}
  (2022) 193} [\href{https://arxiv.org/abs/2111.01161}{{\ttfamily
  2111.01161}}].

\bibitem{vanBeekveld:2022zhl}
M.~van Beekveld, S.~Ferrario~Ravasio, G.~P. Salam, A.~Soto-Ontoso, G.~Soyez and
  R.~Verheyen, \emph{{PanScales parton showers for hadron collisions:
  formulation and fixed-order studies}},
  \href{https://doi.org/10.1007/JHEP11(2022)019}{\emph{JHEP} {\bfseries 11}
  (2022) 019} [\href{https://arxiv.org/abs/2205.02237}{{\ttfamily
  2205.02237}}].

\bibitem{vanBeekveld:2022ukn}
M.~van Beekveld, S.~Ferrario~Ravasio, K.~Hamilton, G.~P. Salam, A.~Soto-Ontoso,
  G.~Soyez et~al., \emph{{PanScales showers for hadron collisions: all-order
  validation}}, \href{https://doi.org/10.1007/JHEP11(2022)020}{\emph{JHEP}
  {\bfseries 11} (2022) 020}
  [\href{https://arxiv.org/abs/2207.09467}{{\ttfamily 2207.09467}}].

\bibitem{Herren:2022jej}
F.~Herren, S.~H\"oche, F.~Krauss, D.~Reichelt and M.~Schoenherr, \emph{{A new
  approach to color-coherent parton evolution}},
  \href{https://doi.org/10.1007/JHEP10(2023)091}{\emph{JHEP} {\bfseries 10}
  (2023) 091} [\href{https://arxiv.org/abs/2208.06057}{{\ttfamily
  2208.06057}}].

\bibitem{vanBeekveld:2023chs}
M.~van Beekveld and S.~Ferrario~Ravasio, \emph{{Next-to-leading-logarithmic
  PanScales showers for Deep Inelastic Scattering and Vector Boson Fusion}},
  \href{https://doi.org/10.1007/JHEP02(2024)001}{\emph{JHEP} {\bfseries 02}
  (2024) 001} [\href{https://arxiv.org/abs/2305.08645}{{\ttfamily
  2305.08645}}].

\bibitem{Preuss:2024vyu}
C.~T. Preuss, \emph{{A partitioned dipole-antenna shower with improved
  transverse recoil}},
  \href{https://doi.org/10.1007/JHEP07(2024)161}{\emph{JHEP} {\bfseries 07}
  (2024) 161} [\href{https://arxiv.org/abs/2403.19452}{{\ttfamily
  2403.19452}}].

\bibitem{Hoche:2024dee}
S.~H\"oche, F.~Krauss and D.~Reichelt, \emph{{The Alaric parton shower for
  hadron colliders}},  \href{https://arxiv.org/abs/2404.14360}{{\ttfamily
  2404.14360}}.

\bibitem{FerrarioRavasio:2023kyg}
S.~Ferrario~Ravasio, K.~Hamilton, A.~Karlberg, G.~P. Salam, L.~Scyboz and
  G.~Soyez, \emph{{Parton Showering with Higher Logarithmic Accuracy for Soft
  Emissions}},
  \href{https://doi.org/10.1103/PhysRevLett.131.161906}{\emph{Phys. Rev. Lett.}
  {\bfseries 131} (2023) 161906}
  [\href{https://arxiv.org/abs/2307.11142}{{\ttfamily 2307.11142}}].

\bibitem{vanBeekveld:2024wws}
M.~van Beekveld et~al., \emph{{A new standard for the logarithmic accuracy of
  parton showers}},  \href{https://arxiv.org/abs/2406.02661}{{\ttfamily
  2406.02661}}.

\bibitem{Beneke:1998ui}
M.~Beneke, \emph{{Renormalons}},
  \href{https://doi.org/10.1016/S0370-1573(98)00130-6}{\emph{Phys. Rept.}
  {\bfseries 317} (1999) 1}
  [\href{https://arxiv.org/abs/hep-ph/9807443}{{\ttfamily hep-ph/9807443}}].

\bibitem{Campbell:2022qmc}
J.~M. Campbell et~al., \emph{{Event generators for high-energy physics
  experiments}},
  \href{https://doi.org/10.21468/SciPostPhys.16.5.130}{\emph{SciPost Phys.}
  {\bfseries 16} (2024) 130}
  [\href{https://arxiv.org/abs/2203.11110}{{\ttfamily 2203.11110}}].

\bibitem{Webber:1997zj}
B.~R. Webber, \emph{{Renormalon phenomena in jets and hard processes}},
  \href{https://doi.org/10.1016/S0920-5632(98)00325-9}{\emph{Nucl. Phys. B
  Proc. Suppl.} {\bfseries 71} (1999) 66}
  [\href{https://arxiv.org/abs/hep-ph/9712236}{{\ttfamily hep-ph/9712236}}].

\bibitem{Dokshitzer_1995}
Y.~Dokshitzer and B.~Webber, \emph{Calculation of power corrections to hadronic
  event shapes},
  \href{https://doi.org/10.1016/0370-2693(95)00548-y}{\emph{Physics Letters B}
  {\bfseries 352} (1995) 451}.

\bibitem{Dokshitzer_1996}
Y.~Dokshitzer, G.~Marchesini and B.~Webber, \emph{Dispersive approach to
  power-behaved contributions in {QCD} hard processes},
  \href{https://doi.org/10.1016/0550-3213(96)00155-1}{\emph{Nuclear Physics B}
  {\bfseries 469} (1996) 93}.

\bibitem{Dokshitzer:1997ew}
Y.~L. Dokshitzer and B.~R. Webber, \emph{{Power corrections to event shape
  distributions}},
  \href{https://doi.org/10.1016/S0370-2693(97)00573-X}{\emph{Phys. Lett. B}
  {\bfseries 404} (1997) 321}
  [\href{https://arxiv.org/abs/hep-ph/9704298}{{\ttfamily hep-ph/9704298}}].

\bibitem{Dokshitzer_1998}
Y.~Dokshitzer, A.~Lucenti, G.~Marchesini and G.~Salam, \emph{Universality of
  corrections to jet-shape observables rescued},
  \href{https://doi.org/10.1016/s0550-3213(97)00650-0}{\emph{Nuclear Physics B}
  {\bfseries 511} (1998) 396}.

\bibitem{Dokshitzer_1998_2}
Y.~L. Dokshitzer, A.~Lucenti, G.~Marchesini and G.~P. Salam, \emph{On the
  universality of the milan factor for 1/{Q} power corrections to jet shapes},
  \href{https://doi.org/10.1088/1126-6708/1998/05/003}{\emph{Journal of High
  Energy Physics} {\bfseries 1998} (1998) 003}.

\bibitem{Dokshitzer:1998qp}
Y.~L. Dokshitzer, G.~Marchesini and G.~P. Salam, \emph{{Revisiting
  nonperturbative effects in the jet broadenings}},
  \href{https://doi.org/10.1007/s1010599c0003}{\emph{Eur. Phys. J. direct}
  {\bfseries 1} (1999) 3}
  [\href{https://arxiv.org/abs/hep-ph/9812487}{{\ttfamily hep-ph/9812487}}].

\bibitem{Banfi:2000si}
A.~Banfi, G.~Marchesini, Y.~L. Dokshitzer and G.~Zanderighi, \emph{{QCD
  analysis of near-to-planar three jet events}},
  \href{https://doi.org/10.1088/1126-6708/2000/07/002}{\emph{JHEP} {\bfseries
  07} (2000) 002} [\href{https://arxiv.org/abs/hep-ph/0004027}{{\ttfamily
  hep-ph/0004027}}].

\bibitem{Banfi:2000ut}
A.~Banfi, Y.~L. Dokshitzer, G.~Marchesini and G.~Zanderighi,
  \emph{{Near-to-planar three jet events in and beyond QCD perturbation
  theory}}, \href{https://doi.org/10.1016/S0370-2693(01)00310-0}{\emph{Phys.
  Lett. B} {\bfseries 508} (2001) 269}
  [\href{https://arxiv.org/abs/hep-ph/0010267}{{\ttfamily hep-ph/0010267}}].

\bibitem{Banfi:2001sp}
A.~Banfi, Y.~L. Dokshitzer, G.~Marchesini and G.~Zanderighi,
  \emph{{Nonperturbative QCD analysis of near - to - planar three jet events}},
  \href{https://doi.org/10.1088/1126-6708/2001/03/007}{\emph{JHEP} {\bfseries
  03} (2001) 007} [\href{https://arxiv.org/abs/hep-ph/0101205}{{\ttfamily
  hep-ph/0101205}}].

\bibitem{Banfi:2001pb}
A.~Banfi, Y.~L. Dokshitzer, G.~Marchesini and G.~Zanderighi, \emph{{QCD
  analysis of D parameter in near to planar three jet events}},
  \href{https://doi.org/10.1088/1126-6708/2001/05/040}{\emph{JHEP} {\bfseries
  05} (2001) 040} [\href{https://arxiv.org/abs/hep-ph/0104162}{{\ttfamily
  hep-ph/0104162}}].

\bibitem{Korchemsky:1999kt}
G.~P. Korchemsky and G.~F. Sterman, \emph{{Power corrections to event shapes
  and factorization}},
  \href{https://doi.org/10.1016/S0550-3213(99)00308-9}{\emph{Nucl. Phys. B}
  {\bfseries 555} (1999) 335}
  [\href{https://arxiv.org/abs/hep-ph/9902341}{{\ttfamily hep-ph/9902341}}].

\bibitem{Korchemsky:2000kp}
G.~P. Korchemsky and S.~Tafat, \emph{{On power corrections to the event shape
  distributions in QCD}},
  \href{https://doi.org/10.1088/1126-6708/2000/10/010}{\emph{JHEP} {\bfseries
  10} (2000) 010} [\href{https://arxiv.org/abs/hep-ph/0007005}{{\ttfamily
  hep-ph/0007005}}].

\bibitem{Bauer:2003di}
C.~W. Bauer, C.~Lee, A.~V. Manohar and M.~B. Wise, \emph{{Enhanced
  nonperturbative effects in Z decays to hadrons}},
  \href{https://doi.org/10.1103/PhysRevD.70.034014}{\emph{Phys. Rev. D}
  {\bfseries 70} (2004) 034014}
  [\href{https://arxiv.org/abs/hep-ph/0309278}{{\ttfamily hep-ph/0309278}}].

\bibitem{Lee:2006nr}
C.~Lee and G.~F. Sterman, \emph{{Momentum Flow Correlations from Event Shapes:
  Factorized Soft Gluons and Soft-Collinear Effective Theory}},
  \href{https://doi.org/10.1103/PhysRevD.75.014022}{\emph{Phys. Rev. D}
  {\bfseries 75} (2007) 014022}
  [\href{https://arxiv.org/abs/hep-ph/0611061}{{\ttfamily hep-ph/0611061}}].

\bibitem{Hoang:2015hka}
A.~H. Hoang, D.~W. Kolodrubetz, V.~Mateu and I.~W. Stewart, \emph{{Precise
  determination of $\alpha_s$ from the $C$-parameter distribution}},
  \href{https://doi.org/10.1103/PhysRevD.91.094018}{\emph{Phys. Rev. D}
  {\bfseries 91} (2015) 094018}
  [\href{https://arxiv.org/abs/1501.04111}{{\ttfamily 1501.04111}}].

\bibitem{Jones:2003yv}
R.~W.~L. Jones, M.~Ford, G.~P. Salam, H.~Stenzel and D.~Wicke,
  \emph{{Theoretical uncertainties on alpha(s) from event shape variables in e+
  e- annihilations}},
  \href{https://doi.org/10.1088/1126-6708/2003/12/007}{\emph{JHEP} {\bfseries
  12} (2003) 007} [\href{https://arxiv.org/abs/hep-ph/0312016}{{\ttfamily
  hep-ph/0312016}}].

\bibitem{Pahl:2009aa}
C.~Pahl, S.~Bethke, O.~Biebel, S.~Kluth and J.~Schieck, \emph{{Tests of
  analytical hadronisation models using event shape moments in $e^{+}e^{-}$
  annihilation}},
  \href{https://doi.org/10.1140/epjc/s10052-009-1167-z}{\emph{Eur. Phys. J. C}
  {\bfseries 64} (2009) 533} [\href{https://arxiv.org/abs/0904.0786}{{\ttfamily
  0904.0786}}].

\bibitem{Davison:2009wzs}
R.~A. Davison and B.~R. Webber, \emph{{Non-Perturbative Contribution to the
  Thrust Distribution in $e^{+} e^{-}$ Annihilation}},
  \href{https://doi.org/10.1140/epjc/s10052-008-0836-7}{\emph{Eur. Phys. J. C}
  {\bfseries 59} (2009) 13} [\href{https://arxiv.org/abs/0809.3326}{{\ttfamily
  0809.3326}}].

\bibitem{Gehrmann:2010uax}
T.~Gehrmann, M.~Jaquier and G.~Luisoni, \emph{{Hadronization effects in event
  shape moments}},
  \href{https://doi.org/10.1140/epjc/s10052-010-1288-4}{\emph{Eur. Phys. J. C}
  {\bfseries 67} (2010) 57} [\href{https://arxiv.org/abs/0911.2422}{{\ttfamily
  0911.2422}}].

\bibitem{Verbytskyi:2019zhh}
A.~Verbytskyi, A.~Banfi, A.~Kardos, P.~F. Monni, S.~Kluth, G.~Somogyi et~al.,
  \emph{{High precision determination of $\alpha_s$ from a global fit of jet
  rates}}, \href{https://doi.org/10.1007/JHEP08(2019)129}{\emph{JHEP}
  {\bfseries 08} (2019) 129}
  [\href{https://arxiv.org/abs/1902.08158}{{\ttfamily 1902.08158}}].

\bibitem{Bethke:2009ehn}
{\scshape JADE} collaboration, S.~Bethke, S.~Kluth, C.~Pahl and J.~Schieck,
  \emph{{Determination of the Strong Coupling alpha(s) from hadronic Event
  Shapes with O(alpha**3(s)) and resummed QCD predictions using JADE Data}},
  \href{https://doi.org/10.1140/epjc/s10052-009-1149-1}{\emph{Eur. Phys. J. C}
  {\bfseries 64} (2009) 351} [\href{https://arxiv.org/abs/0810.1389}{{\ttfamily
  0810.1389}}].

\bibitem{Dissertori:2009ik}
G.~Dissertori, A.~Gehrmann-De~Ridder, T.~Gehrmann, E.~W.~N. Glover,
  G.~Heinrich, G.~Luisoni et~al., \emph{{Determination of the strong coupling
  constant using matched NNLO+NLLA predictions for hadronic event shapes in
  $e^+e^-$ annihilations}},
  \href{https://doi.org/10.1088/1126-6708/2009/08/036}{\emph{JHEP} {\bfseries
  08} (2009) 036} [\href{https://arxiv.org/abs/0906.3436}{{\ttfamily
  0906.3436}}].

\bibitem{OPAL:2011aa}
{\scshape OPAL} collaboration, G.~Abbiendi et~al., \emph{{Determination of
  $alpha_s$ using OPAL hadronic event shapes at $\sqrt{s}=91$ - 209 GeV and
  resummed NNLO calculations}},
  \href{https://doi.org/10.1140/epjc/s10052-011-1733-z}{\emph{Eur. Phys. J. C}
  {\bfseries 71} (2011) 1733}
  [\href{https://arxiv.org/abs/1101.1470}{{\ttfamily 1101.1470}}].

\bibitem{Kardos:2018kqj}
A.~Kardos, S.~Kluth, G.~Somogyi, Z.~Tulip\'ant and A.~Verbytskyi,
  \emph{{Precise determination of $\alpha _{S}(M_Z)$ from a global fit of
  energy\textendash{}energy correlation to NNLO+NNLL predictions}},
  \href{https://doi.org/10.1140/epjc/s10052-018-5963-1}{\emph{Eur. Phys. J. C}
  {\bfseries 78} (2018) 498}
  [\href{https://arxiv.org/abs/1804.09146}{{\ttfamily 1804.09146}}].

\bibitem{dEnterria:2022hzv}
D.~d'Enterria et~al., \emph{{The strong coupling constant: state of the art and
  the decade ahead}}, \href{https://doi.org/10.1088/1361-6471/ad1a78}{\emph{J.
  Phys. G} {\bfseries 51} (2024) 090501}
  [\href{https://arxiv.org/abs/2203.08271}{{\ttfamily 2203.08271}}].

\bibitem{Nason:2023asn}
P.~Nason and G.~Zanderighi, \emph{{Fits of \ensuremath{\alpha}$_{s}$ using
  power corrections in the three-jet region}},
  \href{https://doi.org/10.1007/JHEP06(2023)058}{\emph{JHEP} {\bfseries 06}
  (2023) 058} [\href{https://arxiv.org/abs/2301.03607}{{\ttfamily
  2301.03607}}].

\bibitem{Dokshitzer:1999ai}
Y.~L. Dokshitzer, \emph{{Perturbative QCD and power corrections}},  in
  \emph{{11th Rencontres de Blois on Frontiers of Matter}}, 6, 1999,
  \href{https://arxiv.org/abs/hep-ph/9911299}{{\ttfamily hep-ph/9911299}}.

\bibitem{Dasgupta:1998xt}
M.~Dasgupta and B.~R. Webber, \emph{{Two loop enhancement factor for 1 / Q
  corrections to event shapes in deep inelastic scattering}},
  \href{https://doi.org/10.1088/1126-6708/1998/10/001}{\emph{JHEP} {\bfseries
  10} (1998) 001} [\href{https://arxiv.org/abs/hep-ph/9809247}{{\ttfamily
  hep-ph/9809247}}].

\bibitem{Dasgupta:1999mb}
M.~Dasgupta, L.~Magnea and G.~Smye, \emph{{Universality of 1/Q corrections
  revisited}}, \href{https://doi.org/10.1088/1126-6708/1999/11/025}{\emph{JHEP}
  {\bfseries 11} (1999) 025}
  [\href{https://arxiv.org/abs/hep-ph/9911316}{{\ttfamily hep-ph/9911316}}].

\bibitem{Nason:1995np}
P.~Nason and M.~H. Seymour, \emph{{Infrared renormalons and power suppressed
  effects in $e^{+} e^{-}$ jet events}},
  \href{https://doi.org/10.1016/0550-3213(95)00461-Z}{\emph{Nucl. Phys. B}
  {\bfseries 454} (1995) 291}
  [\href{https://arxiv.org/abs/hep-ph/9506317}{{\ttfamily hep-ph/9506317}}].

\bibitem{Luisoni:2020efy}
G.~Luisoni, P.~F. Monni and G.~P. Salam, \emph{{$C$-parameter hadronisation in
  the symmetric 3-jet limit and impact on $\alpha_s$ fits}},
  \href{https://doi.org/10.1140/epjc/s10052-021-08941-z}{\emph{Eur. Phys. J. C}
  {\bfseries 81} (2021) 158}
  [\href{https://arxiv.org/abs/2012.00622}{{\ttfamily 2012.00622}}].

\bibitem{Caola:2021kzt}
F.~Caola, S.~Ferrario~Ravasio, G.~Limatola, K.~Melnikov and P.~Nason, \emph{{On
  linear power corrections in certain collider observables}},
  \href{https://doi.org/10.1007/JHEP01(2022)093}{\emph{JHEP} {\bfseries 01}
  (2022) 093} [\href{https://arxiv.org/abs/2108.08897}{{\ttfamily
  2108.08897}}].

\bibitem{Caola:2022vea}
F.~Caola, S.~Ferrario~Ravasio, G.~Limatola, K.~Melnikov, P.~Nason and M.~A.
  Ozcelik, \emph{{Linear power corrections to e$^{+}$e$^{-}$ shape variables in
  the three-jet region}},
  \href{https://doi.org/10.1007/JHEP12(2022)062}{\emph{JHEP} {\bfseries 12}
  (2022) 062} [\href{https://arxiv.org/abs/2204.02247}{{\ttfamily
  2204.02247}}].

\bibitem{Dasgupta:2001sh}
M.~Dasgupta and G.~P. Salam, \emph{{Resummation of nonglobal QCD observables}},
  \href{https://doi.org/10.1016/S0370-2693(01)00725-0}{\emph{Phys. Lett. B}
  {\bfseries 512} (2001) 323}
  [\href{https://arxiv.org/abs/hep-ph/0104277}{{\ttfamily hep-ph/0104277}}].

\bibitem{Dasgupta:2002bw}
M.~Dasgupta and G.~P. Salam, \emph{{Accounting for coherence in interjet E(t)
  flow: A Case study}},
  \href{https://doi.org/10.1088/1126-6708/2002/03/017}{\emph{JHEP} {\bfseries
  03} (2002) 017} [\href{https://arxiv.org/abs/hep-ph/0203009}{{\ttfamily
  hep-ph/0203009}}].

\bibitem{Salam:2001bd}
G.~P. Salam and D.~Wicke, \emph{{Hadron masses and power corrections to event
  shapes}}, \href{https://doi.org/10.1088/1126-6708/2001/05/061}{\emph{JHEP}
  {\bfseries 05} (2001) 061}
  [\href{https://arxiv.org/abs/hep-ph/0102343}{{\ttfamily hep-ph/0102343}}].

\bibitem{Lee:2024esz}
K.~Lee, A.~Pathak, I.~Stewart and Z.~Sun, \emph{{Nonperturbative Effects in
  Energy Correlators: From Characterizing Confinement Transition to Improving
  $\alpha_s$ Extraction}},  \href{https://arxiv.org/abs/2405.19396}{{\ttfamily
  2405.19396}}.

\bibitem{Chen:2024nyc}
H.~Chen, P.~F. Monni, Z.~Xu and H.~X. Zhu, \emph{{Scaling violation in power
  corrections to energy correlators from the light-ray OPE}},
  \href{https://arxiv.org/abs/2406.06668}{{\ttfamily 2406.06668}}.

\bibitem{Dasgupta:2003iq}
M.~Dasgupta and G.~P. Salam, \emph{{Event shapes in $e^+ e^-$ annihilation and
  deep inelastic scattering}},
  \href{https://doi.org/10.1088/0954-3899/30/5/R01}{\emph{J. Phys. G}
  {\bfseries 30} (2004) R143}
  [\href{https://arxiv.org/abs/hep-ph/0312283}{{\ttfamily hep-ph/0312283}}].

\bibitem{Catani:1996vz}
S.~Catani and M.~H. Seymour, \emph{{A General algorithm for calculating jet
  cross-sections in NLO QCD}},
  \href{https://doi.org/10.1016/S0550-3213(96)00589-5}{\emph{Nucl. Phys. B}
  {\bfseries 485} (1997) 291}
  [\href{https://arxiv.org/abs/hep-ph/9605323}{{\ttfamily hep-ph/9605323}}].

\bibitem{DMO}
Y.~L. Dokshitzer, G.~Marchesini and G.~Oriani, \emph{{Measuring color flows in
  hard processes: Beyond leading order}},
  \href{https://doi.org/10.1016/0550-3213(92)90211-S}{\emph{Nucl. Phys. B}
  {\bfseries 387} (1992) 675}.

\bibitem{Catani:1999ss}
S.~Catani and M.~Grazzini, \emph{{Infrared factorization of tree level QCD
  amplitudes at the next-to-next-to-leading order and beyond}},
  \href{https://doi.org/10.1016/S0550-3213(99)00778-6}{\emph{Nucl. Phys. B}
  {\bfseries 570} (2000) 287}
  [\href{https://arxiv.org/abs/hep-ph/9908523}{{\ttfamily hep-ph/9908523}}].

\bibitem{gavin}
F.~Caola, C.~Farren-Colloty, J.~Helliwell, R.~Patel, G.~P. Salam and S.~Zanoli,
  \emph{{private communication}}, {\emph{paper in preparation} (2024) }.

\bibitem{Hatta:2013iba}
Y.~Hatta and T.~Ueda, \emph{{Resummation of non-global logarithms at finite
  $N_c$}}, \href{https://doi.org/10.1016/j.nuclphysb.2013.06.021}{\emph{Nucl.
  Phys. B} {\bfseries 874} (2013) 808}
  [\href{https://arxiv.org/abs/1304.6930}{{\ttfamily 1304.6930}}].

\bibitem{Dasgupta:2007wa}
M.~Dasgupta, L.~Magnea and G.~P. Salam, \emph{{Non-perturbative QCD effects in
  jets at hadron colliders}},
  \href{https://doi.org/10.1088/1126-6708/2008/02/055}{\emph{JHEP} {\bfseries
  02} (2008) 055} [\href{https://arxiv.org/abs/0712.3014}{{\ttfamily
  0712.3014}}].

\bibitem{Dasgupta:2013ihk}
M.~Dasgupta, A.~Fregoso, S.~Marzani and G.~P. Salam, \emph{{Towards an
  understanding of jet substructure}},
  \href{https://doi.org/10.1007/JHEP09(2013)029}{\emph{JHEP} {\bfseries 09}
  (2013) 029} [\href{https://arxiv.org/abs/1307.0007}{{\ttfamily 1307.0007}}].

\bibitem{Hoang:2019ceu}
A.~H. Hoang, S.~Mantry, A.~Pathak and I.~W. Stewart, \emph{{Nonperturbative
  Corrections to Soft Drop Jet Mass}},
  \href{https://doi.org/10.1007/JHEP12(2019)002}{\emph{JHEP} {\bfseries 12}
  (2019) 002} [\href{https://arxiv.org/abs/1906.11843}{{\ttfamily
  1906.11843}}].

\bibitem{Dasgupta:2009tm}
M.~Dasgupta and Y.~Delenda, \emph{{On the universality of hadronisation
  corrections to QCD jets}},
  \href{https://doi.org/10.1088/1126-6708/2009/07/004}{\emph{JHEP} {\bfseries
  07} (2009) 004} [\href{https://arxiv.org/abs/0903.2187}{{\ttfamily
  0903.2187}}].

\bibitem{Hounat:2023nvc}
F.~Hounat, \emph{{Non-universal Milan factors for QCD jets}},
  \href{https://doi.org/10.1007/JHEP06(2024)022}{\emph{JHEP} {\bfseries 06}
  (2024) 022} [\href{https://arxiv.org/abs/2311.06089}{{\ttfamily
  2311.06089}}].

\bibitem{Cacciari:2008gp}
M.~Cacciari, G.~P. Salam and G.~Soyez, \emph{{The anti-$k_t$ jet clustering
  algorithm}}, \href{https://doi.org/10.1088/1126-6708/2008/04/063}{\emph{JHEP}
  {\bfseries 04} (2008) 063} [\href{https://arxiv.org/abs/0802.1189}{{\ttfamily
  0802.1189}}].

\bibitem{Dasgupta:2016bnd}
M.~Dasgupta, F.~A. Dreyer, G.~P. Salam and G.~Soyez, \emph{{Inclusive jet
  spectrum for small-radius jets}},
  \href{https://doi.org/10.1007/JHEP06(2016)057}{\emph{JHEP} {\bfseries 06}
  (2016) 057} [\href{https://arxiv.org/abs/1602.01110}{{\ttfamily
  1602.01110}}].

\bibitem{Hoang:2024zwl}
A.~H. Hoang, O.~L. Jin, S.~Pl\"atzer and D.~Samitz, \emph{{Matching
  Hadronization and Perturbative Evolution: The Cluster Model in Light of
  Infrared Shower Cutoff Dependence}},
  \href{https://arxiv.org/abs/2404.09856}{{\ttfamily 2404.09856}}.

\bibitem{Korchemsky:1997sy}
G.~P. Korchemsky, G.~Oderda and G.~F. Sterman, \emph{{Power corrections and
  nonlocal operators}}, \href{https://doi.org/10.1063/1.53732}{\emph{AIP Conf.
  Proc.} {\bfseries 407} (1997) 988}
  [\href{https://arxiv.org/abs/hep-ph/9708346}{{\ttfamily hep-ph/9708346}}].

\end{thebibliography}\endgroup

\end{document}